\documentclass[11pt,a4paper]{article}

\usepackage[T1]{fontenc}
\usepackage[utf8]{inputenc}
\usepackage{lmodern}
\usepackage{microtype}
\usepackage{tabularx,booktabs,seqsplit,siunitx,cite}
\usepackage{threeparttable}
\usepackage{array} 

\usepackage{amsmath,amssymb,amsthm,bm,mathtools}

\newtheorem{theorem}{Theorem}[section]

\theoremstyle{definition}
\newtheorem{definition}[theorem]{Definition}

\usepackage[margin=2.4cm]{geometry}
\usepackage{setspace}
\usepackage{authblk}
\usepackage{titlesec}
\titleformat{\section}{\normalfont\large\bfseries}{\thesection}{0.6em}{}
\titleformat{\subsection}{\normalfont\bfseries}{\thesubsection}{0.6em}{}
\usepackage{graphicx}
\usepackage{xcolor}
\usepackage{subcaption}

\definecolor{Navy}{RGB}{0,47,95}
\definecolor{Gold}{RGB}{255,204,0}
\definecolor{LBlue}{RGB}{235,245,255}
\definecolor{LGold}{RGB}{255,247,214}
\definecolor{Steel}{RGB}{52,120,180}
\definecolor{Orange}{RGB}{220,120,40}
\definecolor{DkGray}{RGB}{90,90,90}

\definecolor{doneGreen}{HTML}{2E7D32}
\definecolor{ongoOrange}{HTML}{E65100}
\definecolor{todoGray}{HTML}{616161}

\usepackage{tikz}
\usetikzlibrary{arrows.meta,positioning,calc,shapes.geometric,fit,backgrounds}

\usepackage{booktabs}
\usepackage{array}
\usepackage{multirow}
\usepackage{caption}

\usepackage{enumitem}
\setlist[itemize,1]{leftmargin=1.5em,itemsep=2pt,topsep=2pt}
\setlist[enumerate,1]{leftmargin=1.8em,itemsep=2pt,topsep=2pt}

\usepackage[colorlinks=true,linkcolor=blue!60!black,
            citecolor=blue!60!black,urlcolor=blue!60!black]{hyperref}
\usepackage[capitalize,noabbrev]{cleveref}
\newif\ifincludeSI
\includeSItrue   

\newcommand{\Prb}{\mathbb{P}}

\title{Uncovering higher-order interactions from memory in observed dynamics}

\author[1,2]{Arthur Rondeau\thanks{Correspondence: \texttt{arthur.rondeau@unige.ch}}}
\author[3]{Alain Barrat}
\author[1,2]{Roland Bouffanais}
\affil[1]{\small Department of Computer Science, University of Geneva, Switzerland}
\affil[2]{\small Global Studies Institute, University of Geneva, Switzerland}
\affil[3]{\small Aix Marseille Univ, Universit\'e de Toulon, CNRS, CPT, Marseille, France}
\date{}

\begin{document}
\bibliographystyle{unsrt}
\maketitle

\begin{abstract}
  Many complex systems involve group interactions, which can readily be represented by higher-order structures such as hypergraphs. 
  However, when observing a system's dynamical evolution, establishing whether that evolution actually results from higher-order interactions, or whether pairwise interactions suffice to explain the data, remains a challenge.
  Even in the simple case of a temporal sequence of nodes visited during a diffusion process, such a task is intricate.
  Indeed, group interactions leave their traces as memory on the observed nodes, while path dependence in a purely pairwise process can mimic that same footprint without any group structure.
  To resolve this, we propose an inference pipeline based on a hidden Markov model (HMM) that tests, directly from an observed sequence of nodes carrying no group labels, whether higher-order interactions are required to describe the dynamics.
  The HMM latent states represent higher-order interactions, seen as hyperedges of an unobserved hypergraph,
  so that their detection becomes a model selection problem 
  between the HMM class and a class of null models without higher-order interactions.
    We first validate this approach on synthetic node sequences generated from random walks on graphs and hypergraphs, showing that it recovers both the number and the composition of the underlying hyperedges, while correctly rejecting them when the generating process
  is pairwise. As a second step, we consider a validation based on empirical data, namely empirical trajectories of ball possession in football matches.
  This application shows that the pipeline goes beyond providing statistical evidence of the presence of higher-order interactions, and yields interpretable underlying structures: the recovered hyperedges map directly onto recognizable tactical team formations.
  \end{abstract}

\section{Introduction}
\label{sec:intro}
Systems involving higher-order interactions have recently attracted significant attention~\cite{battiston2020,Battiston2021,bick_what_2023,iacopini_temporal_2024},
with the aim of going beyond the representation of complex systems as sets of pairwise interactions between their components.
In particular, hypergraphs have become widely used for representing collective phenomena in which the relevant unit of interaction is a group of elements, represented
as a hyperedge.
These representations have been adopted in many different settings, such as scientific collaborations through 
co-authorship networks~\cite{nr}, ecological systems \cite{mayfield_higher-order_2017, bairey_high-order_2016} or social group dynamics~\cite{iacopini_temporal_2024}.

Two main approaches have emerged in the literature concerned with higher-order interactions. 
The first is the forward problem perspective, which takes the group structure as given, and studies the impact of such interactions on dynamical processes such as contagion~\cite{iacopini2019,battiston2020,malizia_hyperedge_2025, arruda_contagion_2024, matamalas_abrupt_2020}, synchronisation \cite{skardal_higher_2020, malizia_hyperedge_2025}, or consensus formation \cite{neuhauser_consensus_2021, neuhauser_multi-body_2019}.
A second line of research addresses the inverse problem and seeks to infer a higher-order structure from observed node dynamics, either by characterising higher-order statistical dependencies in multivariate time series~\cite{santoro_higher-order_2023}, or by reconstructing the hypergraph that generates the dynamics of interacting units time-series data~\cite{delabays_hypergraph_2025}.

While these inference methods provide valuable insights into higher-order interactions, the evidence they provide is often indirect: a higher-order structure is typically deemed present when it improves the statistical fit to the observed data, yet improved fit alone does not guarantee that the process is effectively generated by a higher-order mechanism. 
This key limitation is effectively an identifiability problem: whichever model class is granted more flexibility will fit the observed data at least as well, so a better fit under the assumption of group interactions cannot, on its own, be read as evidence for them. 
The comparison between model classes becomes informative only when the complexity of each model class is explicitly penalised~\cite{peixoto_graphs_2026,arenas_when_2026,rommens_beyond_2026}, and when the higher-order model is benchmarked against pairwise alternatives afforded with comparable resources in terms of memory, latent variables and parameters~\cite{villegas_meloni_2026}.
This gap has recently been identified as a central open problem for the field~\cite{peixoto_graphs_2026}, and the conditions under which such a comparison is meaningful have been detailed~\cite{rommens_beyond_2026,villegas_meloni_2026,arenas_when_2026}.
Only then is a preference for a higher-order representation grounded in a principled, falsifiable claim rather than an artifact of flexibility.

Let us consider here the specific case in which the data at hand is a temporal sequence of visited nodes, resulting from a diffusive process on the network---i.e., a random walk following the interactions between nodes.
Interestingly, if the interactions at play are higher-order, a statistical signature of these group dependencies emerges in the sequence as temporal memory.
This observation is tied to the fact that the term higher-order has been used in the literature to describe two distinct properties. On the one hand, 
temporal correlations in observed paths call for representations beyond first-order Markov dynamics~\cite{Rosvall2014,lambiotte2019}. 
These higher-order Markov chain methods successfully capture non-Markovianity and rigorously test it against lower-order dependencies, but they do so by expanding the state space of an otherwise pairwise process, without assuming any multi-body group interactions. 
On the other hand, the term higher-order has more recently come to refer to  actual group interactions, often represented using hyperedges and hypergraphs~\cite{battiston2020,Battiston2021}. 
 However, as mentioned above, inference techniques in this second sense typically reconstruct structures without controlling for temporal correlations, thus leaving open whether a memory-driven pairwise process could explain the same observations.
Establishing a genuine structural origin for the observed dependencies therefore requires formalising this distinction: a higher-order generative explanation must be tested directly against parsimonious alternatives, specifically memoryless and memory-based pairwise scenarios, to determine which best explains the data. This defines the central research question we address:
Given only an observed process, here a discrete sequence of nodes, is a higher-order representation required to account for the interactions between nodes generating this dynamics, or do pairwise interactions, with or without memory, describe the data at least as parsimonously?

To address this problem, we propose a framework that bridges the two perspectives mentioned above. 
We posit that these two notions are in fact linked: when the group interaction active at a given time influences which group interaction follows, the sequence of nodes exhibits temporal memory that can serve as the observable trace of the underlying hyperedges.
Conversely, a purely pairwise and memoryless process leaves no such trace.
Building on this, we introduce a hidden Markov model in which the latent states, which carry the temporal memory, explicitly represent the higher-order structures themselves, namely the hyperedges. 
Deciding whether a higher-order representation is genuinely required thus becomes a concrete, testable question, which we answer through model selection against memoryless and memory-based pairwise baselines. 
Crucially, this selection relies not on in-sample fit alone but on out-of-sample fit together with an explicit penalty on model complexity~\cite{villegas_meloni_2026,peixoto_graphs_2026}. When the observed dynamics exhibit memory, a higher-order process may be at play, and model comparison then tests, rather than assumes, that the selected representation is the one the data support.

We first validate this approach on synthetic node sequences generated by random walks on hypergraphs, including negative controls in which the sequences are generated by memoryless or memory-based pairwise processes.
We then apply the framework to an empirical data set of the same form,  namely trajectories of ball possession in football matches from the StatsBomb open dataset~\cite{statsbomb_opendata}.
In both cases, our method not only tests whether higher-order interactions are required to account for the observed process, but also returns the underlying structure itself, the set of hyperedges. These identified hyperedges recover the ground truth in the synthetic case and can be meaningfully interpreted in the light of domain knowledge in the empirical one. 
Specifically, for the football empirical data, the recovered latent states map---i.e., the hyperedges---map onto recognisable tactical team formations, with no supervision on team arrangement.

Overall, our contribution is twofold. First, we introduce a hyperedge-latent hidden Markov model that bridges higher-order Markov processes and higher-order structure reconstruction. By recasting the relevance of higher-order representations as a complexity-penalised model selection problem against pairwise baselines, it provides a principled , falsifiable test of whether such a representation is genuinely required to account for an observed process.
Second, we show that the recovered structures are interpretable when the framework is applied to real-world empirical data: on football possession sequences, the inferred hyperedges map onto tactical team formations, thus giving the statistical result a direct domain-knowledge reading.

\section{Results}
\label{sec:results}

\subsection{Problem Statement}

Let us consider a set of temporal trajectories of varying length over a vertex set 
$V$ of size $|V|$.
Each trajectory is thus a sequence of visited vertices $(v_t)_{t\in\mathbb{N}}$, with $v_t\in V$, of the kind a random walk could generate; such a sequence constitutes an observed process. Beyond the trajectories themselves, no
information about the mechanism that produced them is assumed, nor about the structure
of interactions connecting the vertices. Given
only these data, we ask whether the dynamics driving the observed
process are shaped by higher-order interactions among groups of
vertices, and, if so, whether the underlying group structure can be recovered.

\subsection{Pipeline Overview}
\label{sec:results_overview}

\begin{figure}[t]
  \centering
  \resizebox{\textwidth}{!}{
  \begin{tikzpicture}[
    >=Stealth,
    font=\footnotesize,
    panel/.style={draw=Navy!28, line width=0.7pt, rounded corners=4pt, fill=Navy!2},
    outcomeOK/.style={draw=Steel, fill=Steel!12, line width=0.9pt, rounded corners=4pt, 
                      align=center, font=\bfseries\small, inner sep=8pt},
    outcomeHi/.style={draw=Orange, fill=Orange!12, line width=0.9pt, rounded corners=4pt, 
                      align=center, font=\bfseries\small, inner sep=8pt},
    obs/.style={circle, draw=Navy, fill=LBlue, line width=0.9pt,
                minimum size=0.55cm, inner sep=0pt, font=\bfseries\tiny},
    smallobs/.style={circle, draw=Navy, fill=LBlue, line width=0.7pt,
                     minimum size=0.45cm, inner sep=0pt, font=\bfseries\tiny},
    smalllat/.style={circle, draw=Orange, fill=LGold, line width=0.7pt,
                     minimum size=0.45cm, inner sep=0pt, font=\bfseries\tiny},
    ptitle/.style={font=\bfseries\small, color=Navy, align=center},
    plabel/.style={font=\bfseries\large, color=Navy},
    flowOK/.style={->, line width=1.4pt, Steel!85!black,
                   shorten >=2pt, shorten <=2pt, rounded corners=6pt},
    flowHigh/.style={->, line width=1.4pt, Orange!85!black,
                     shorten >=2pt, shorten <=2pt, rounded corners=6pt},
    branchLbl/.style={font=\scriptsize\bfseries, fill=white, inner sep=3pt}
  ]


  \node[panel, minimum width=5.6cm, minimum height=4.6cm] (Pa) at (0, 0) {};

  \node[plabel, anchor=north west] at ($(Pa.north west)+(0.15, -0.15)$) {};
  \node[ptitle, anchor=south, font=\large \bf] at ($(Pa.north)+(0, -0.80)$) {Observed Process};

  \node[obs, minimum size=0.8cm, font=\bfseries ] (a1) at ($(Pa.center)+(-1.40, 0.50)$) {$v_1$};
  \node[obs, minimum size=0.8cm, font=\bfseries ] (a2) at ($(Pa.center)+( 0.00, 0.50)$) {$v_2$};
  \node[obs, minimum size=0.8cm, font=\bfseries ] (a3) at ($(Pa.center)+( 1.40, 0.50)$) {$v_3$};
  \node[font=\bfseries, color=Navy] (adots)
    at ($(Pa.center)+(2.50, 0.50)$) {$\cdots$};

  \draw[->, semithick, Navy] (a1) -- (a2);
  \draw[->, semithick, Navy] (a2) -- (a3);
  \draw[->, semithick, Navy] (a3) -- ($(adots.center)+(-0.35, 0)$);

\node[font=\scriptsize\itshape, color=black!60, align=center, font=\bfseries \small]
  at ($(Pa.center)+(0, -0.80)$)
  {Vertex sequence $(v_t)_{t\in\mathbb{N}}$\\no group label};
  
  \node[panel, minimum width=4.6cm, minimum height=4.6cm, right=2.2cm of Pa] (Pb) {};

  \node[plabel, anchor=north west] at ($(Pb.north west)+(0.15, -0.15)$) {};
  \node[ptitle, anchor=south, font=\large \bf] at ($(Pb.north)+(0, -0.80)$) {I. Markovianity Test};

  \begin{scope}[shift={($(Pb.center)+(0, 0.10)$)}]
    \node[font=\large, align=center] at (0, -0.20)
      {$P(v_{t+1} \mid v_t, v_{t-1}, \ldots)$\\[6pt]
       $\overset{?}{=}$\\[6pt]
       $P(v_{t+1} \mid v_t)$};
  \end{scope}

  \node[outcomeOK, minimum width=3.0cm, minimum height=3.6cm, right=3.2cm of Pb,
        font=\bfseries \large, xshift=0.4cm] (pairwise) {Pairwise\\sufficient};


  \node[panel, minimum width=5.6cm, minimum height=5.2cm, below=2.2cm of Pa] (Pc) {};

  \node[plabel, anchor=north west] at ($(Pc.north west)+(0.15, -0.15)$) {};
  \node[ptitle, anchor=south, font=\large\bfseries] at ($(Pc.north)+(0, -0.70)$) {II. Learn and Build HMM};

  \node[smalllat, minimum size=0.75cm, font=\bfseries] (he1) at ($(Pc.center)+(-1.70, 0.85)$) {$e_{t{-}1}$};
  \node[smalllat, minimum size=0.75cm, font=\bfseries] (he2) at ($(Pc.center)+( 0.00, 0.85)$) {$e_t$};
  \node[smalllat, minimum size=0.75cm, font=\bfseries] (he3) at ($(Pc.center)+( 1.70, 0.85)$) {$e_{t{+}1}$};

  \node[smallobs, minimum size=0.75cm, font=\bfseries] (hv1) at ($(Pc.center)+(-1.70, -0.85)$) {$v_{t{-}1}$};
  \node[smallobs, minimum size=0.75cm, font=\bfseries] (hv2) at ($(Pc.center)+( 0.00, -0.85)$) {$v_t$};
  \node[smallobs, minimum size=0.75cm, font=\bfseries] (hv3) at ($(Pc.center)+( 1.70, -0.85)$) {$v_{t{+}1}$};

  \draw[->, thick, Orange] (he1) -- (he2) node[midway, above, font=\scriptsize, yshift=-1pt] {$T$};
  \draw[->, thick, Orange] (he2) -- (he3) node[midway, above, font=\scriptsize, yshift=-1pt] {$T$};
  \draw[->, thick, Navy] (he1) -- (hv1);
  \draw[->, thick, Navy] (he2) -- (hv2) node[midway, right, font=\scriptsize, xshift=0pt] {$B$};
  \draw[->, thick, Navy] (he3) -- (hv3);

  \draw[->, thick, Navy, dashed] (hv1.north east) to[bend right=25]
      node[midway, right, font=\scriptsize\bfseries, xshift=-2pt, yshift=-3pt] {$S$} (he2.south west);
  \draw[->, thick, Navy, dashed] (hv2.north east) to[bend right=25] (he3.south west);

  \node[font=\scriptsize\itshape, color=black!65, align=center, font=\bfseries \small]
      at ($(Pc.center)+(0, -1.95)$)
      {hyperedge-latent HMM};

  \node[panel, minimum width=5.6cm, minimum height=2.7cm,
        right=2.2cm of Pc, yshift=1.2cm, xshift=-0.7cm] (Pd_top) {};

  \node[ptitle, anchor=south,font=\large\bfseries] at ($(Pd_top.north)+(0, -0.7)$)
    {III. Model Comparison};

  \begin{scope}[shift={($(Pd_top.center)+(0, 0.0)$)}]
    \node[font=\large, align=center] at (0, -0.2)
      {HMM against
       null models via\\[4pt]
       Likelihood \& BIC};
  \end{scope}

  \node[panel, minimum width=5.6cm, minimum height=2.3cm,
        below=0.2cm of Pd_top] (Pd_bot) {};

  \node[ptitle, anchor=south,font=\large\bfseries] at ($(Pd_bot.north)+(0, -0.70)$)
    {IV. Interpretation};

  \begin{scope}[shift={($(Pd_bot.center)+(0, 0.0)$)}]
    \node[font=\large, align=center] at (0, -0.2)
      {Hypergraph Representation\\[4pt]
       $\hat{B}_{e,v},\quad \hat{S}_{v,e},\quad \hat{T}_{e,e'}$};
  \end{scope}

\node[outcomeHi, minimum width=2.4cm, minimum height=5.2cm, right=2.2cm of Pd_top,
      font=\bfseries \large, xshift=1.5cm, yshift=-1.2cm] (higherorder) {Higher-\\order\\relevant};

  \draw[flowOK] (Pa.east) -- (Pb.west);

  \draw[flowOK] (Pb.east) -- (pairwise.west)
      node[midway, branchLbl, text=Steel!85!black, yshift=8pt,
          font=\bfseries\normalsize] {If Markovian};

  \draw[flowHigh] (Pb.south) -- ++(0,-0.8) coordinate (bend1) -| (Pc.north);
  \path (Pb.south |- bend1) -- (Pc.north |- bend1)
      node[midway, branchLbl, text=Orange!90!black, yshift=8pt,
          font=\bfseries\normalsize] {If non-Markovian};

  \draw[flowOK] (Pc.east |- Pd_top.west) -- (Pd_top.west);

  \draw[flowHigh] (Pd_top.east) -- (higherorder.west |- Pd_top.east)
      node[midway, above=2pt, align=center, branchLbl, text=Orange!90!black,
           font=\bfseries\small] {HMM outperforms\\null models};

  \draw[flowHigh] (higherorder.west |- Pd_bot.east) -- (Pd_bot.east);

  \draw[flowOK] (Pd_top.north) -- ++(0, 1.3) coordinate (bend2) -| (pairwise.south);
  \path (Pd_top.north |- bend2) -- (pairwise.south |- bend2)
      node[midway, branchLbl, text=Steel!85!black, yshift=8pt,
          font=\bfseries\normalsize] {HMM not better than null models};

  \end{tikzpicture}
  } 
\caption{\textbf{Overview of the analysis pipeline.}
From an observed vertex sequence $(v_t)_{t\in\mathbb{N}}$ lacking any group label, 
\textbf{I.}~a conditional-mutual-information test checks whether the process is Markovian; if so, pairwise interactions suffice. 
\textbf{II.}~Otherwise, a hidden Markov model (HMM) whose latent states are hyperedges is fitted: vertices are emitted by the current hyperedge ($B$), 
the process evolves from one hyperedge to the next ($T$), and each 
selected hyperedge depends on the previous vertex ($S$, dashed). 
\textbf{III.}~The HMM is compared with null pairwise models through held-out likelihood and Bayesian information criterion (BIC); if it outperforms them, higher-order interactions are relevant. 
\textbf{IV.}~The fitted $\hat{B}$, $\hat{S}$ and $\hat{T}$ then give an interpretable hypergraph representation of the structure and process.}
  \label{fig:pipeline}
\end{figure}
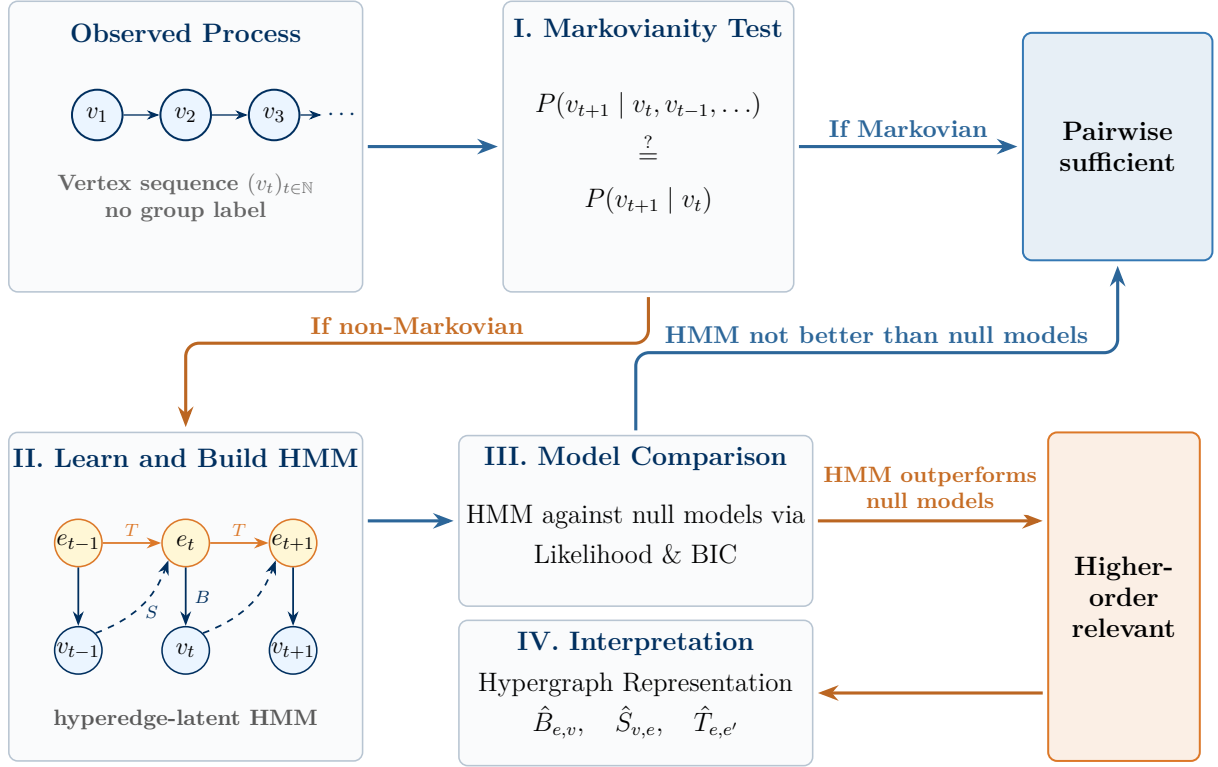

Intuitively, the premise underlying our approach is as follows. If the process generating an observed sequence of nodes over time is governed by pairwise interactions (i.e., the next vertex depends only on the current vertex), then it is Markovian on $V$. If, instead, it is governed by higher-order interactions (typically represented by a hypergraph), then conditioning on $V$ alone is not sufficient: the process exhibits memory, appearing non-Markovian on $V$, while remaining Markovian on the augmented state space $V\times E$, where $E$ is the set of hyperedges. Based on this premise, we propose a four-stage pipeline, schematically depicted in Figure~\ref{fig:pipeline}.

\subsubsection{Step I. Markovianity Test}
First, given an observed process, we perform a non-Markovianity test (Methods~\ref{sec:methods_cmi}). This acts as a natural initial filter: if the sequence is Markovian, a network representing pairwise interactions between the nodes is sufficient to describe the dynamics. 
However, if the test confirms non-Markovianity, it indicates that pairwise interactions on $V$ are not enough information. This motivates the next step to investigate whether this memory actually comes from structural higher-order interactions.

\subsubsection{Step II. Hidden Markov Model}
\label{sec:HMM_presentation}
To recover these higher-order interactions 
from the observed process (sequence of nodes over time), we introduce below a hidden Markov model (HMM) in which the latent variables represent the (unobserved) hyperedges (the set $E$). This approach builds upon an extensive literature on hypergraph random walks \cite{chitra2019,carletti_random_2020,carletti_random_2021,eriksson_how_2021}. However, while existing models successfully capture structural features such as non-reversibility, they remain almost exclusively memoryless (Markovian) on the vertex set: when the walker is on a given
vertex, the next visited hyperedge depends only on the current vertex. 
Here, we go beyond this paradigm by equipping the hyperedge-selection process with a temporal memory. By embedding memory at the latent level, the observed trajectory appears non-Markovian on $V$, while remaining strictly Markovian on the augmented state space $V\times E$. Let us now describe the HMM in detail.

Intuitively, the dynamics can be pictured as a random walker evolving on a hypergraph through a coupled, two-step mechanism (Figure~\ref{fig:pipeline}.II). Assuming the walker is currently at vertex $v_t$, to which it arrived via hyperedge $e_t$ (hence $v_t \in e_t$):
\begin{enumerate}
    \item \textbf{Latent Transition:} A candidate hyperedge $e_{t+1}^{\mathrm{cand}}$ is proposed. This "proposal" is only based on a transition probability matrix between hyperedges, independently from the current vertex $v_t$.
    \item \textbf{Acceptance \& Emission:} This candidate is accepted with a probability proportional to its relevance to the current vertex $v_t$: this "relevance" encodes in particular the topology of the hypergraph (in particular, $e_{t+1}$ needs to include $v_t$). Once the new hyperedge $e_{t+1}$ is selected, the walker moves to the next vertex $v_{t+1} \in e_{t+1}$ according to an internal distribution of node weights within that hyperedge.
\end{enumerate}
\vspace{0.5cm}
The output of this mechanism is thus a sequence of visited vertices over time, with the visited hyperedges serving only as transitory, unobserved steps.

More formally, this process takes place over the $|V|$ observed vertices and along $K$ hidden states (representing the hyperedges of the unobserved interaction structure), and 
is fully specified by three row-stochastic matrices. The first two are:
\begin{definition}[HMM Matrices]
\  \label{def:hmm_matrices}
  \begin{itemize}
      \item \textbf{Affinity} $T \in \mathbb{R}^{K \times K}$: $T_{e,e'}$ is the transition probability from hidden state $e$ to subsequent hidden state $e'$, conditionally independent of the vertices currently visited. It is the primary carrier of \textbf{hidden memory}.
      \item \textbf{Assignment/Acceptance} $S \in \mathbb{R}^{|V| \times K}$: $S_{v,e}$ is the probability of selecting hidden state (hyperedge) $e$ given the current vertex $v$.
  \end{itemize}
\end{definition}
In this framework, $T$ thus encodes the affinity between latent unobserved hyperedges (the proposal step in the intuitive description above), while $S$ anchors that affinity to the actual vertex being visited at time $t$ (the acceptance step):
for a walker on a vertex $v$ ($\in e$) at time $t$,
the hyperedge $e'$ at time $t+1$ is selected according to
\begin{equation}
    \varphi(v, e', e)
    = \frac{T_{e, e'} \, S_{v, e'}}
           {\sum_{\tilde{e} \in E} T_{e, \tilde{e}} \, S_{v, \tilde{e}}}.
    \label{eq:phi_kernel_intro}
\end{equation}
The third row-stochastic matrix 
$B$ dictates the selection of the next observed vertex at time $t+1$:
\begin{definition}[HMM Matrix B]
\ \label{def:hmm_matrices_B}
  \begin{itemize}
      \item \textbf{Emission} $B \in \mathbb{R}^{K \times |V|}$: $B_{e,v}$ is the probability of selecting the vertex $v$ within the hidden state $e$.
  \end{itemize}
\end{definition}

Model parameters are fit by maximizing the conditional likelihood of the observed process, $P(v_2,\dots,v_{\ell} \mid S,B,T,v_1)$, where $\ell$ is the trajectory length (Methods~\ref{sec:methods_hmm},~\ref{sec:methods_fitting}).
Note that the number of hidden states $K$ is treated as a free parameter, since the true number of hyperedges $|E|$ is a priori unknown: we rely on model selection to  recover the 
dimension $K=|E|$ of the underlying system.

\subsubsection{Step III. Model Comparison}
In addition to fitting the HMM model to the data at hand (i.e., to a set of vertex trajectories of various lengths), we benchmark the HMM against three baselines using held-out likelihood and the Bayesian Information Criterion (BIC), a complexity penalization metric, in order to ensure the inferred structure is meaningful.

First, we compare it to an HMM where $T$ is uniform: the selection of the next latent state
depends thus only on the matrix $S$, i.e., on which hyperedges contain the current
vertex $v_t$. This removes all latent memory and reduces the model to a generalisation of 
the edge-dependent vertex-weighted (EDVW) random walk of Chitra and 
Raphael~\cite{chitra2019}. In their model, a walker at vertex $v$ first 
selects an incident hyperedge $e$ according to hyperedge weights, then moves 
to a vertex $v'$ of $e$ according to edge-dependent vertex weights. These two 
steps correspond to our matrices $S$ (hyperedge selection) and $B$ (vertex 
selection within the hyperedge), so that the observed process is a 
first-order Markov chain on $V$ with transition matrix 
$P_{\mathrm{SB}} = SB$. In~\cite{chitra2019}, $S$ and $B^\top$ are 
constrained to share the same support, namely the incidence structure of the 
hypergraph, with weights fixed by the data. In our setting, this constraint 
is relaxed: $S$ and $B$ are free row-stochastic matrices learned from the 
trajectories, and their supports need not coincide.
This serves as our memoryless hypergraph null model (SB model); any predictive gain over it comes specifically from the latent temporal structure encoded in $T$. 

We also compare the HMM against standard first-order (MC1) and second-order 
(MC2) Markov chains on the vertex set $V$. These baselines discard 
the hypergraph structure and model the dynamics directly on observed 
vertices: MC1 is parametrised by a $|V| \times |V|$ transition matrix, while MC2 
conditions on the pair of previous vertices $(v_{t-1}, v_t)$ through a 
$|V|^2 \times |V|$ matrix, and thus captures a single step of explicit 
higher-order memory.
The comparison with MC2 acts as a safeguard: it ensures that any detected non-Markovianity actually reflects a latent hyperedge structure, rather than just ordinary temporal memory on the nodes.
For details on baseline definition, fitting and evaluation, see 
Methods~\ref{sec:methods_fitting}, \ref{sec:methods_nulls} and 
\ref{sec:methods_comparison}.

\subsubsection{Step IV. Interpretation of Higher-Order Structure}

If the model selection procedure selects HMM as the best one to explain the data at hand,
we use the learned matrices to describe and interpret the higher-order structure 
of interactions between the nodes governing the dynamics.
In particular, the assignment ($S$) and emission ($B$) matrices jointly reveal the topology of the latent hypergraph, indicating which vertices are grouped together to form hyperedges, while
the transition matrix ($T$) describes the dynamics unfolding between those hyperedges.

\subsection{Validation on Synthetic Hypergraphs}
\label{sec:synthetic_results}

To validate our pipeline, we first consider a controlled setting where the ground-truth generative process and higher-order underlying structure are known. 
Specifically, we consider a homogeneous Watts--Strogatz hypergraph \cite{raghav_smallworldness_2023} with 
$|V|= 15$ nodes and $|E|=15$ hyperedges, and
we generate vertex trajectories (i.e. random walks trajectories on the hypergraph) 
directly from the two-steps HMM mechanism described previously in  
Section~\ref{sec:HMM_presentation}, with known matrices $T$, 
$S$ and $B$ as described in Methods ~\ref{sec:methods_synthetic}.
This yields trajectories that are, by construction, non-Markovian on $V$ but Markovian on 
$V\times E$.

\begin{figure*}[h!]
  \centering
  \includegraphics[width=\textwidth]{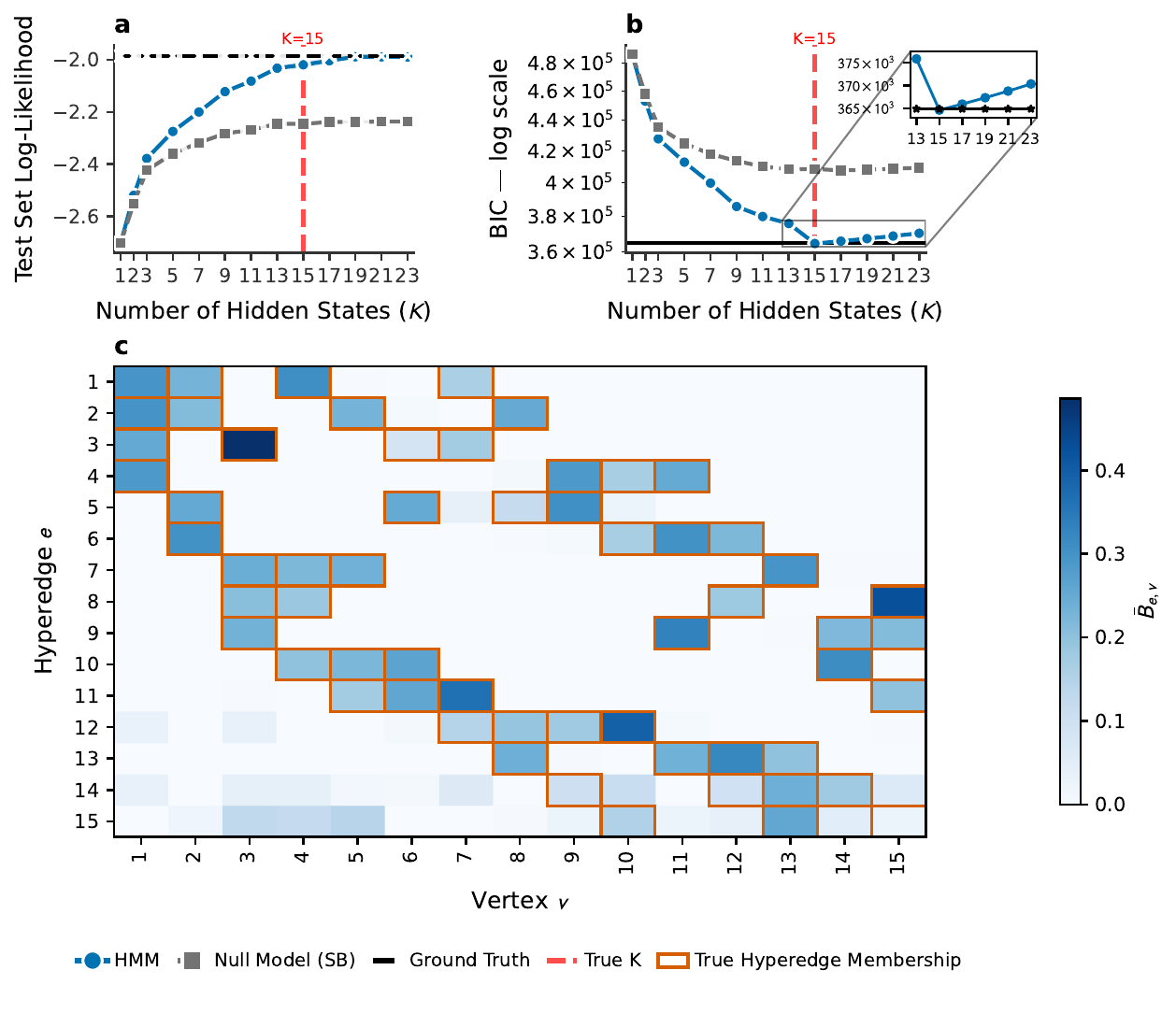}
  \caption{\textbf{Model performance and recovered parameters on synthetic data}
  ($|E|=15$ hyperedges of size 4, $|V|=15$ vertices, Watts–Strogatz rewiring
  probability $p=10^{-3}$, and overlap parameter $l=2$ controlling the degree
  of vertex sharing between neighbouring hyperedges).
  \textbf{a}, Test-set log-likelihood per step as a function of the number of
  hidden states $K$, comparing the HMM against the SB null model and ground
  truth. Shaded bands show $\pm 1$ standard deviation across 10
  cross-validation folds. The dashed red line marks the true number of
  hidden states.
  \textbf{b}, Bayesian Information Criterion (BIC, log scale) across models
  as a function of $K$; lower is better.
  \textbf{c}, Fold-averaged emission matrix $\bar{B}_{e,v}$ at the true
  $K$; rows are hierarchically clustered (Ward linkage) and aligned to the
  ground-truth incidence matrix; vermilion
  boxes mark the ground-truth hyperedge
  support.}
  \label{fig:synth-homo-reports-and-params-R_normal}
\end{figure*}

Figure~\ref{fig:synth-homo-reports-and-params-R_normal} summarizes results for the likelihood, complexity based tests when varying $K$.
As expected, the edge-latent HMM reaches the best performance against all baselines on both 
predictive likelihood and complexity-penalized comparison (Table \ref{tab:model_comparison_k15_p_0.35_R_normal}), approaching the 
ground-truth log-likelihood as $K$ increases
(Fig.~\ref{fig:synth-homo-reports-and-params-R_normal}a), and reaching the 
lowest BIC across the explored range of number of latent state  
(Fig.~\ref{fig:synth-homo-reports-and-params-R_normal}b). Notably, the minimum BIC is reached when the number of hidden states matches the ground truth one ($K=15$): the model recovers the intrinsic complexity of the system without supervision. 
It is however worth noting that in general, 
the HMM's peak performance (highest likelihood or 
lowest BIC) does not necessarily occur exactly at the ground-truth $K$; here 
for instance, the log-likelihood curve plateaus slightly later. 
This is because, unlike the baseline models, which admit a closed-form solution, the HMM parameters are optimized via expectation-maximization coupled with 
gradient descent, an optimization landscape that can shift the observed 
plateau by a few units of $K$ (for more details see Methods~\ref{sec:methods_fitting}). The exact location of this plateau should therefore be interpreted with some caution, especially when no ground truth value is available (as in the case of empirical data).

The recovered parameters of the HMM confirm the fidelity of this estimation. 
Figure~\ref{fig:synth-homo-reports-and-params-R_normal}c)
shows the recovered hypergraph structure at $K=15$ 
(corresponding to the ground truth number of hyperedges)
as encoded in the learned emission matrix $\hat{B}_{e,v}$. $\hat{B}$ 
reproduces the block structure of the ground-truth incidence matrix describing
which nodes belong to which hyperedges (vermilion boxes), with each recovered 
hyperedge concentrating its probability mass on the correct set of member 
vertices (Fig.~\ref{fig:synth-homo-reports-and-params-R_normal}c). 

We also show the learned matrix $\hat{S}$ in the Supplementary 
Figure~\ref{fig:synth_homo_panel1_SI}: $S$ and $B$ share the same support in the ground truth (i.e., $B_{e,v} = S_{v,e} = 0$ when $v \notin e$), and the HMM correctly learns 
$\hat{S}$ with a
support similar to $\hat{B}$'s.
Likewise, the recovered transition matrix $\hat{T}_{e,e'}$ closely matches the ground truth, with a cosine similarity of $0.910 \pm 0.050$ (mean $\pm$ s.d.\ across the $k=10$ folds; $0.918 \pm 0.041$ for $\hat{B}$ and $0.892 \pm 0.021$ for $\hat{S}$).

Together, these results show that the model correctly recovers both the composition of each hyperedge and also the temporal dynamics coupling them in the actual 
process used to generate the synthetic data.

Moreover, while the true value of $K$ is here known, this would not be the case in a realistic
situation. Given the potential unvertainty on the value of $K$ highlighted above, it is thus important to investigate how the inferred higher-order structure
depends on $K$. We show in \ref{sec:Supplementary-homogeneous} how the structure evolves when $K$
increases, and in particular that the recovered structure is robust, additional
hyperedges largely being redundant copies of hyperedges already found at lower $K$
(see Supplementary Tables~\ref{tab:cross_K_refinement} and~\ref{tab:within_K_redundancy}).

We also show in the Supplementary material results for a different ground truth distribution of $B$ and $S$, as well as on a heterogeneous hypergraph (Supplementary Table~\ref{tab:model_comparison_k15_p_0.35}, Supplementary Figure~\ref{fig:synth-homo-reports-and-params},
~\ref{sec:appendix_synthetic_heterogeneous}), confirming the
robustness of our results with respect to the synthetic structure considered. 
We further find that the hypergraph representation remains relevant even at low magnitude of the higher order interactions (\ref{sec:appendix_homogeneous_power}).
Finally, we examine how hyperedge size and overlap between hyperedges affect the advantage of 
the higher-order representation with respect to a pairwise baseline in \ref{sec:appendix_synthetic_overlap_size}.

For completeness, we consider as well
two negative controls experiments (\ref{sec:appendix_negative_synth}),
using synthetic node sequences generated without higher-order interactions, by 
a purely Markovian process on the one hand, and by a temporal second-order process
on the other hand. On Markovian data, the HMM does not perform better 
than the SB null model
(Supplementary Table \ref{tab:model_comparison_k15_p_0}); on 
data generated by a second-order process, MC2  
outperforms the HMM (Supplementary Table \ref{tab:model_comparison_k15_mc2}). Together, these checks support the central claim: when 
the HMM outperforms all competing models, the detected non-Markovianity can 
be legitimately attributed to higher-order interactions rather than to ordinary temporal memory.

\begin{table}[htbp]
  \centering
  \begin{threeparttable}
    \caption{
    Model comparison for homogeneous hypergraph with $p_{stay}=0.35$ at true $K$ ($K=15$) and at its largest
    BIC-dominant $K=23$. Test-set log-likelihood (mean $\pm$ 1 s.d.
    across $n=10$ cross-validation folds), BIC, and the difference in
    test log-likelihood relative to MC2, the strongest
    baseline.}
  \label{tab:model_comparison_k15_p_0.35_R_normal}
  \setlength{\tabcolsep}{27pt}
  \begin{tabular}{l c c c}
  \toprule
  \textbf{Model} & \textbf{Test LL (nats)} & \textbf{BIC} & {$\Delta$} \\
  \midrule
  Null Model (SB) & $-2.246 \pm 0.005$ & \num{407545} & -0.207 \\
  MC1 & $-2.237 \pm 0.006$ & \num{404251} & -0.198 \\
  MC2 & $-2.039 \pm 0.007$ & \num{399037} & ref. \\
  \textbf{HMM ($K=15$)} & $-2.019 \pm 0.014$ & \textbf{\num{364595}} & +0.020** \\
  HMM ($K=23$)$^\dagger$ & $\mathbf{-1.989 \pm 0.006}$ & \num{370342} & +0.050*** \\
  \midrule
  \textit{Ground Truth} & $-1.985 \pm 0.006$ & \num{364846} & -- \\
  \bottomrule
  \end{tabular}
  \begin{tablenotes}\footnotesize
  \item[$\dagger$] $K=23$ is the largest $K$ at which the HMM's BIC (computed on a single fit to the full dataset) is lower than or equal to every baseline's full-dataset BIC. This is a descriptive comparison of point estimates, not a significance claim.
  \item[*/**/***] $p<0.05/0.01/0.001$ (one-sided Nadeau--Bengio corrected paired $t$-test on test log-likelihood; $n=10$ folds).
  \item[$\Delta$] Difference in mean test log-likelihood relative to MC2, the strongest baseline (positive favors the row's model). This is the only valid basis for a significance claim, since it is computed on paired, fold-matched differences; the mean $\pm$ 1 s.d. reported per row describes each model's own fold-to-fold variability and should not be compared for overlap. BIC values are point estimates from a single full-dataset fit and are not subject to this test.
  \item[] Ground truth is shown as an oracle reference (upper bound on achievable performance) and is not subject to a significance test.
  \end{tablenotes}
  \end{threeparttable}
  \end{table}


\subsection{Empirical Results on Football Formations}
\label{sec:empirical_results}

Let us now consider actual empirical data sets for which our pipeline can readily
be applied: empirical data on ball possessions trajectories. We leverage
the StatsBomb data \cite{statsbomb_opendata} set as described in Methods~\ref{sec:methods_empirical}:
each trajectory consists in a sequence of player roles of a given team
receiving and owning the ball during a game. The ball indeed goes from player to player
through passes, yielding temporal sequences of player roles such as the two following
examples:
\begin{quote}
  \noindent\textit{Trajectory 1:}\\
  \texttt{Right Center Back} $\rightarrow$ \texttt{Left Defensive Midfield} $\rightarrow$ \texttt{Right Defensive Midfield;}
  
  \medskip
  \noindent\textit{Trajectory 2:}\\
  \texttt{Right Center Back} $\rightarrow$ \texttt{Right Back} $\rightarrow$ \texttt{Right Center Back.}
\end{quote}

These trajectories are dictated by the team's strategy and we hypothesize that the 
trajectories result from a set of interactions between players that might involve
group interactions and not only pairwise ones. In particular, we hypothesize that the
choice of the next player receiving the ball is 
not entirely dictated by the previous one (first order Markov process) or even the two previous (second order Markov process) but also by the tactical groups formed by the players.
Our hypothesis is thus that the announced tactical formation used during the match
might correspond to higher-order structures that can be unveiled by our analysis pipeline.

In particular, we present here results for possession trajectories aggregated from 
two professional football teams for different announced formations: Bayern Munich (formations 3-4-2-1 / 3-5-1-1) and Atlético Madrid (formation 4-4-2).
Note that the observed nodes of the process are not the players themselves but rather
player roles, and the announced formation (i.e., the label given by \textit{StatsBomb} team) corresponds to the expected higher-order interactions. 
Importantly, a single player may occupy multiple roles across different phases of play. 
There is therefore no \textit{necessary} 
equivalence between a formation (sets of players) and the hyperedges (set of players roles) recovered by our model. 
Nevertheless, for a given announced formation (or a coherent set of formations), 
we expect that certain roles should 
co-occur within the same hyperedges, while others should typically not, reflecting the 
tactical structure the formation encodes.

We first checked for Markovianity in all data sets considered. In all cases, the trajectories passed the conditional mutual information test for non-Markovianity 
(Supplementary Table~\ref{tab:cmi}), establishing that a first-order Markov 
description is insufficient and that a higher-order latent structure is a legitimate modelling target. Let us now consider separately the outcome of our analysis
on the two teams.

\subsubsection{Bayern Munich (Formations (3-4-2-1) and (3-5-1-1))}

\begin{figure}[htbp]
  \centering
  \includegraphics[width=\textwidth]{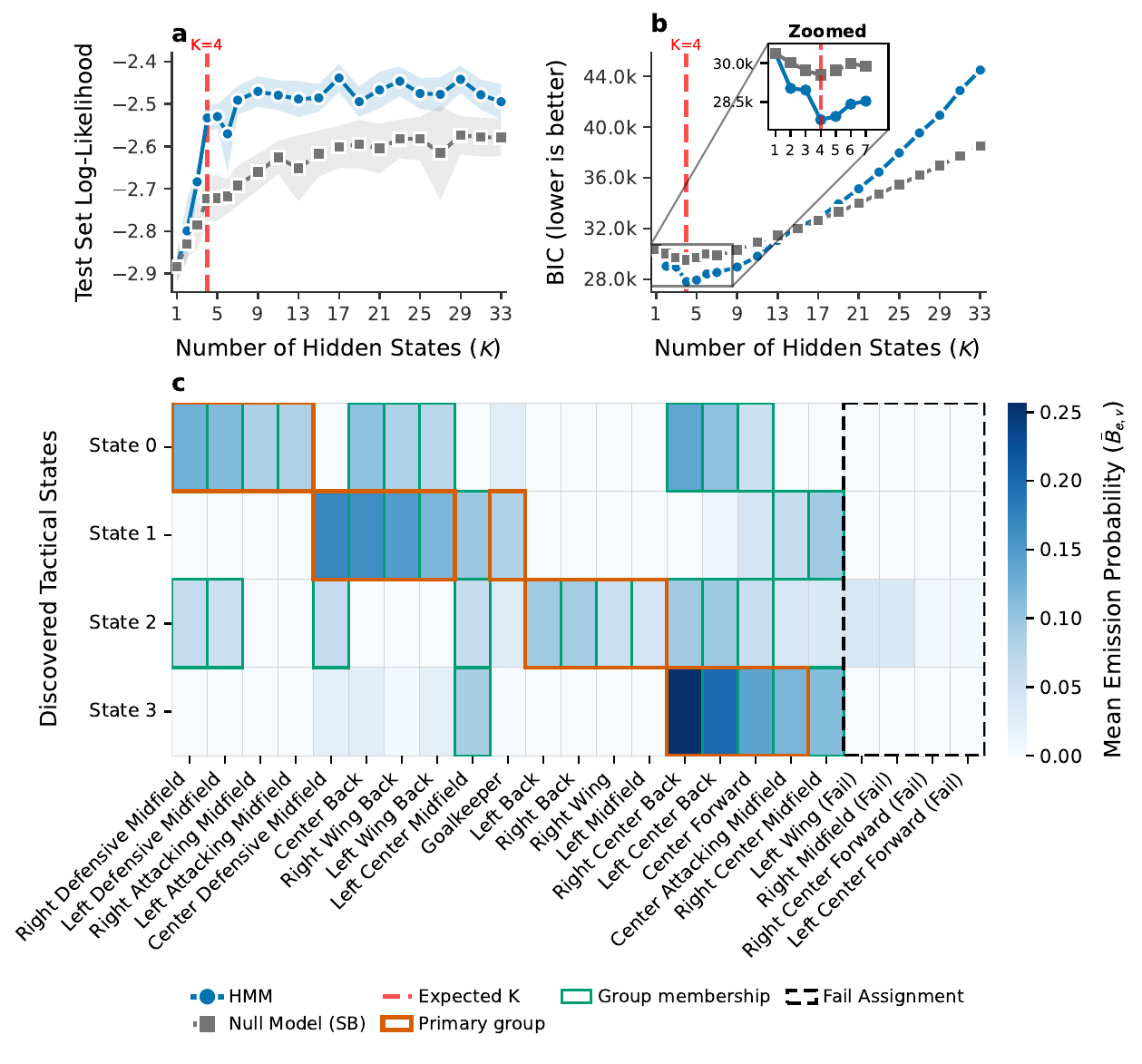}
  \caption{\textbf{Model performance and recovered tactical states for
  Bayern Munich} (formations 3-4-2-1 / 3-5-1-1, expected $K=4$).
  \textbf{a}, Test-set log-likelihood per step as a function of the number
  of hidden states $K$, comparing the HMM against the SB null model. The
  dashed red line marks the expected number of hidden states.
  \textbf{b}, Bayesian Information Criterion (BIC, log scale) across models
  as a function of $K$; lower is better.
  \textbf{c}, Fold-averaged emission matrix $\bar{B}_{e,v}$ at the expected
  $K$. Each role is assigned to its most probable discovered tactical
  state (highest mean emission probability across cross-validation
  folds). Significance is assessed via a standard $t$-test on the
  fold-to-fold empirical variability, at the 95\% confidence level.
  Vermillion boxes mark roles for which the top state is both
  significantly above chance and significantly more probable than the
  role's next most likely state, defining the composition of each
  recovered hyperedge. Green boxes mark roles significantly above chance
  for a given state but without a single dominant state, indicating
  meaningful association with more than one state. Roles not significant
  above chance for any state are collected under the dashed black box
  and labelled \textit{Fail}.}
  \label{fig:bayern}
\end{figure}

Unlike in the synthetic case, here we do not know the true higher order structure, 
albeit the announced formation gives us hints about it.
The BIC curve of the HMM allows us to identify the characteristic
higher-order size of the system, i.e., the effective number of hyperedges. 
Figure \ref{fig:bayern}b illustrates this in the Bayern Munich example, for which the $K$
corresponding to the lowest BIC matches the announced formations
size ($K=4$), recovering the tactical complexity of the formation without supervision.
The HMM achieves the best performance, with a higher held-out log-likelihood and a lower BIC than all baselines (Table~\ref{tab:model_comparison_k4_bayern}), at both $K=4$ and $K=9$. 
These correspond, respectively, to the expected number of formations (the announced size, with the lowest BIC) and to the ceiling $K$ that we define as the highest $K$ where the HMM still achieves a lower BIC than the baselines. 
As in the synthetic setting, this indicates that the latent hyperedge structure captures predictive signal beyond what pairwise transitions alone can explain.

To interpret the higher-order structure emerging from the analysis
pipeline, we rely on the estimated matrices $\hat{B}$ and $\hat{T}$,
as in the case of the synthetic data (Section~\ref{sec:synthetic_results}).
The matrix $\hat{B}$ is shown in Figure~\ref{fig:bayern}c at the BIC-optimal $K$,
describing which roles compose each hyperedge. 
It is important to note that here, $\hat{B}$ cannot be compared to any ground truth.
Instead, the figure reports, for every role, how reliably it can be assigned to a
discovered tactical state (hyperedge) across cross-validation folds. Roles with a
clear, statistically robust assignment are visually grouped together,
while roles lacking such an assignment are set apart. Details on this procedure
are given in the figure caption.
Figure~\ref{fig:bayern}c shows that
most discovered hyperedges concentrate their probability mass on a small, tactically coherent subset of roles. In particular, State (hyperedge) 0 recovers the central midfield engine shared by
the 3-4-2-1 and 3-5-1-1 formations, combining the double pivot with
both attacking midfield roles, while State 1 groups the central
defensive midfielder, center back, wing backs, and goalkeeper that
define their shared defensive build.

The matrix $\hat{S}$ is shown in Supplementary Figure~\ref{fig:bayern_munich_panel1_SI}.
Note that, unlike in the synthetic case, there is here no a priori guarantee that 
$S$ and $B$ share the same support. We indeed
find that $\hat{S}$ and $\hat{B}$ share only a partial common support,
indicating an asymmetry in vertex--hyperedge affinities.
Finally, the recovered $\hat{T}$ matrix shown in Figure~\ref{fig:recovered_T_bayern}
allows to analyze the dynamics between the recovered tactical groups:
the central midfield (State 0) and the wide/flank
configuration (State 2) both act as highly stable, self-sustaining phases,
with self-transition probabilities of 0.986 and 0.997, respectively,
indicating that once play settles into either of these shapes, it tends to
persist there over many consecutive frames rather than oscillating rapidly
between configurations. In contrast, what could be called the deep
build-up state (State 1) and the advanced, final-third state (State 3)
form a tightly coupled pair: the combined self- and cross-transition
probability between them accounts for over 99\% of the outgoing
probability mass from each state ($P(1\to1)=0.524$, $P(1\to3)=0.467$;
$P(3\to3)=0.404$, $P(3\to1)=0.585$), with negligible probability of
routing through either State 0 or State 2 (see Supplementary
Figure~\ref{fig:recovered_T_bayern}). Rather than a
gradual possession-progression cycle passing through an intermediate
midfield phase, this pattern suggests a direct, vertical link between the
team's deepest and most advanced structures, consistent with rapid
transitions or direct progression that bypass a settled midfield shape.
Notably, State 3 retains both center backs among its defining vertices,
which may reflect the wide center backs in this back-three system pushing
high to support the attack during these advanced phases. 
Note that this direct coupling between
build-up and final-third configurations emerges from the latent dynamics
without any explicit supervision on possession phase.

\subsubsection{Case of the Atl\'etico Madrid (Formation 4-4-2)}

As already discussed in the case of synthetic data, the value of $K$ obtained
from the optimisation procedure should be taken with some caution. It is thus
crucial to examine how the learned structure evolves with $K$, and
in particular whether that evolution is coherent rather than arbitrary or random.
This is especially important when applying the HMM framework to
empirical data, where the true underlying structure is unknown
and $K$ cannot be chosen with the same confidence as in a controlled or
domain-informed setting. In such cases, a coherent evolution of hyperedges when $K$
is varied can give confidence in the results of the pipeline: 
even if the selected $K$ is not exactly the right one, the
resulting analysis remains then consistent and thus interpretable.

To investigate this point, we focus on the case study of the Atl\'etico Madrid (formation 4-4-2). For this team, the $K$ corresponding to the lowest BIC is $K=9$,
which exceeds the nominal formation size of a 4-4-2; this is however not unexpected,
since, as discussed above, the vertices encode tactical roles
rather than individual players, and a single player can occupy multiple
roles across different phases of play. 

\begin{figure}[htbp]
    \centering
    \includegraphics[width=1\textwidth]{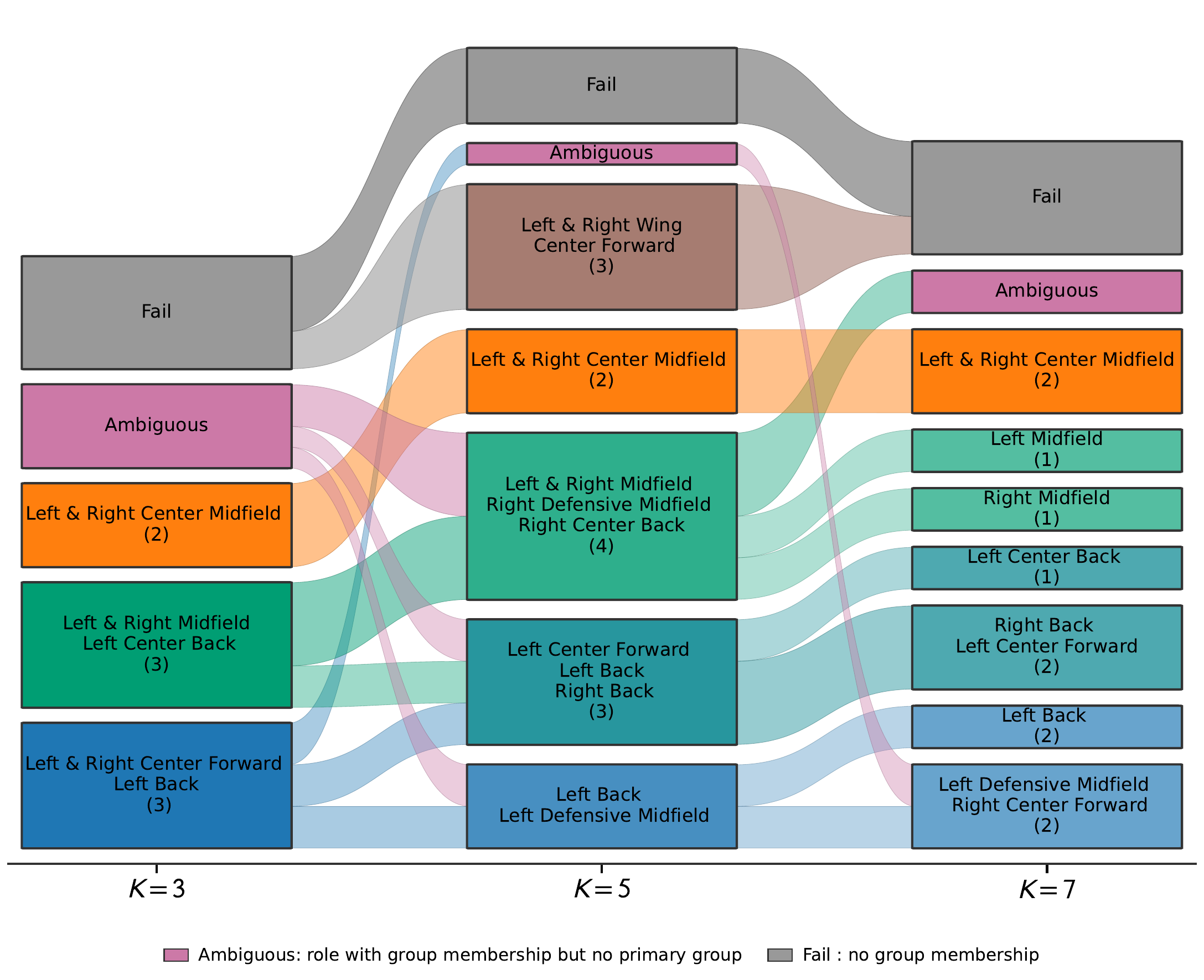}
    \caption{\textbf{Evolution of the learned tactical structure for
    Atl\'etico Madrid (4-4-2) across $K=3$, $K=5$, and $K=7$.} Each column
    shows the discovered tactical states at one value of $K$, with box
    height proportional to the number of vertices (digit in parenthesis) assigned to that state
    (fold-averaged emission matrix $\bar{B}_{e,v}$, argmax assignment).
    Ribbons connect states across adjacent $K$ values, with ribbon width
    proportional to the number of vertices shared between them, showing
    how groupings split, merge, or persist as $K$ increases. Vertices that fail to clear the chance-baseline significance test are
    pooled into the gray \textit{Fail} category at each $K$; vertices that
    clear chance but do not significantly dominate their runner-up state are
    instead labeled \textit{Ambiguous}. Both tests use an ordinary
    (uncorrected) fold-based $t$-test, consistent with the descriptive
    statistics used elsewhere in this pipeline. The \textit{Fail} category's
    height is compressed by a factor of 0.3 relative to its true count for
    legibility, as noted on the figure.}
    \label{fig:atletico_evolution}
\end{figure}

We first note that, even if $K=3$ (the domain-expected value matching the announced 4-4-2) is 
not the optimal choice (cf Supplementary Table~\ref{tab:model_comparison_atletico_madrid_442_k3}), 
the structure learned at $K=3$
remains highly interpretable and closely corresponds to the announced
formation. Tracking how these states reorganize as $K$ increases to $5$
and then $7$ (Fig.~\ref{fig:atletico_evolution})
further shows that this structure does not fragment randomly, but evolves coherently: 
several groupings, such as the central
midfield pairing, are recovered consistently across all three values of
$K$, while others progressively refine into more specific role
combinations. This result supports the idea 
that the recovered structure remains interpretable and stable across
different choices of the number of hidden states, and not only at a single $K$. 
However, regarding the likelihood, it is only at the ceiling $K=11$ (highest $K$ where the HMM still achieves a lower BIC than the baselines) that the HMM attains a positive information gap.
This behavior illustrates a possible trade-off between the predictive
performance attainable at larger $K$ and the interpretability of the
recovered structure, which can become increasingly fragmented at this
scale. For extended results on likelihood and BIC performances, see Supplementary
Figure~\ref{fig:atletico_442_SI}.

\subsubsection{Other teams}

We generalize the results described above to other formations and teams in 
\ref{sec:appendix_mega_plot_formations_logos}. 
In most cases, the edge-latent HMM attains the lowest BIC at the expected $K$ and a positive information gain over the baselines at the ceiling $K$, confirming that the higher-order interactions are consistently detected across a broader range of tactical formations and teams.

Moreover, the results for Bayern Munich and Atlético Madrid are robust to the random seed used in the optimisation procedures: we repeated the full fitting pipeline over several independent seeds, each changing the random initialisation of the model parameters and the partition of the trajectories into training, validation and held-out sets (\ref{sec:appendix_mega_plot_seeds_logos}).

\begin{table}[htbp]
  \centering
  \begin{threeparttable}
  \caption{Model comparison at expected $K$ ($K=4$) and at its largest
  full-dataset BIC-dominant $K=9$. Test-set log-likelihood (mean $\pm$ 1 s.d.
  across $n=10$ cross-validation folds), full-dataset BIC, and the difference in
  test log-likelihood relative to MC1, the strongest
  baseline.}
  \label{tab:model_comparison_k4_bayern}
  \setlength{\tabcolsep}{27pt}
  \begin{tabular}{l c c c}
  \toprule
  \textbf{Model} & \textbf{Test LL (nats)} & \textbf{BIC} & {$\Delta$} \\
  \midrule
  Null Model (SB) & $-2.723 \pm 0.053$ & \num{29543} & -0.165 \\
  MC2 & $-2.663 \pm 0.053$ & \num{123595} & -0.105 \\
  MC1 & $-2.558 \pm 0.033$ & \num{30453} & ref. \\
  \textbf{HMM ($K=4$)} & $-2.533 \pm 0.031$ & \textbf{\num{27814}} & +0.026* \\
  HMM ($K=9$)$^\dagger$ & $\mathbf{-2.470 \pm 0.036}$ & \num{28965} & +0.088*** \\
  \bottomrule
  \end{tabular}
  \begin{tablenotes}\footnotesize
  \item[$\dagger$] $K=9$ is the largest $K$ at which the HMM's BIC is lower than or equal to every baseline's full-dataset BIC (i.e. ceiling $K$).
  \item[*/**/***] $p<0.05/0.01/0.001$ (one-sided Nadeau--Bengio corrected paired $t$-test on test log-likelihood; $n=10$ folds).
  \item[$\Delta$] Difference in mean test log-likelihood relative to MC1, the strongest baseline (positive favors the row's model). This is the only valid basis for a significance claim, since it is computed on paired, fold-matched differences; the mean $\pm$ 1 s.d. reported per row describes each model's own fold-to-fold variability and should not be compared for overlap. BIC values are point estimates from a single full-dataset fit and are not subject to this test.
  \end{tablenotes}
  \end{threeparttable}
  \end{table}

\section{Discussion}
\label{sec:discussion}

Taken together, our results provide an answer to the problem stated in the introduction: 
our pipeline is able to determine whether a higher-order representation is actually necessary to describe the process at the origin of the considered data. 
The debiased conditional mutual information
test can first consistently rule out any first-order pairwise explanation of the
observed dynamics, and the hyperedge-latent HMM can be
compared that same pairwise baseline, and with first- and
second-order Markov-chain baselines, on both held-out likelihood and
complexity-penalised model comparison. When the HMM is selected through model comparison,
the observed non Markovianity is not merely evidence that a
higher-order process could be at play, it is the observable trace
that lets us test, rather than assume, that it is.

Beyond the illustration of the pipeline on synthetic data resulting from either
pairwise or higher-order interactions, application on empirical data confirm the 
ability of the pipeline not only to test for higher-order but to uncover interpretable structures.
Across multiple football tactical configurations from
different teams, latent tactical phases indeed emerge from possession
sequences, the held-out likelihood advantage over pairwise and
memory-expanded baselines is significant, and per-role emission profiles
align with recognizable tactical units. Notably,
in the case where the announced formation and the BIC-optimal number of
hidden states (i.e., of hyperedges) 
coincide (e.g. Bayern Munich), the model recovers the team's
formation directly from possession sequences alone, with no supervision
on team shape.

This agreement between domain expectation and statistical optimum is not
universal, however, and this divergence is itself informative. For
Atl\'etico Madrid, the domain-expected number of hyperedges $K$ does not correspond to the
BIC-optimal solution, which instead lies at higher $K$. 
Despite this, the representation recovered at
the lower, domain-expected $K$ remains tactically coherent, and tracking how tactical states reorganise as $K$ increases
shows that this coherence is not a coincidence of one particular value of $K$: the
same groupings are recovered consistently across several values of $K$,
In other words, interpretability of the recovered structures 
is robust against uncertainty in the choice of the number $K$ of these higher-order 
interactions: this is a particularly important point as in general for empirical data, 
$K$ might be known a priori only approximately or not at all, and the optimisation procedure
is not guaranteed to provide the exact ground truth value of $K$.

\paragraph{Limitations and outlook.}

A first limitation concerns the class of dynamics considered. Determining 
whether a higher-order representation is actually necessary to describe the 
process at the origin of the observed data is a broad and crucial theoretical 
question, and our pipeline addresses it here only for a restricted class: 
random-walk-like trajectories on a fixed hypergraph. This class is nonetheless 
fundamental, since such walks underlie diffusion processes on networks. 
Moreover, the proposed model comparison framework is not specific to random walks: it provides a natural starting point for other classes of dynamics for which the role of 
higher-order structure has been explored 
\cite{robiglio_synergistic_2025,santoro_higher-order_2023}, and could offer a 
principled way to assess whether such structure is needed to explain the 
observed dynamics.

A second limitation follows from the hypothesis of a fixed higher-order 
structure assumed in the HMM, and concerns thus also 
the type of data to which our framework can be directly 
applied. In particular, for a given $K$, the HMM learns a single
higher-order representation held constant over time across all
trajectories. The empirical dataset should thus satisfy this property for the
model to be meaningfully applicable. 
Ball trajectories collected in football matches were a good
candidate for this reason: it is reasonable to assume that a team's
tactical formation, i.e. the higher-order structure itself, remains
approximately constant over the course of a match, and therefore over
its pass trajectories.

Related to the scarcity of suitable datasets, a third limitation concerns the 
size of individual datasets. Since the HMM is fit by gradient-based optimisation 
rather than in closed form, the weaker the underlying higher-order signal, the 
more trajectories are required to recover it reliably. For this reason, we 
aggregated several formations for teams without enough trajectories per 
individual formation; Atl\'etico Madrid (formation 4-4-2) is the only exception, 
for which a single formation provided sufficient data on its own.

A natural direction for addressing the scarcity of suitable datasets is to allow
the model to learn a temporally varying higher-order representation,
extending the HMM framework to datasets in which the higher-order
structure itself evolves over time. This would, however,
considerably expand both the computation time and the search space,
while reducing the effective sample size if trajectories are
partitioned into temporal subsets. In a setting where the size of the dataset is not
a limiting factor, this represents a promising direction for future work.

\section{Methods}
\label{sec:methods}

\subsection{Test of Markovianity via Conditional Mutual Information}
\label{sec:methods_cmi}

To test whether trajectories deviate from a first-order Markov process, we 
compute the conditional mutual information~\cite{papapetrou_markov_2013}
\begin{equation}
    I(v_{t+1}; v_{t-1} \mid v_t) 
    = \sum_{a,b,c} P(v_{t-1}{=}a,\, v_t{=}b,\, v_{t+1}{=}c)\,
      \log \frac{P(v_{t+1}{=}c \mid v_{t-1}{=}a,\, v_t{=}b)}
                 {P(v_{t+1}{=}c \mid v_t{=}b)}.
\end{equation}
Under the Markov assumption, $v_{t+1} \perp v_{t-1} \mid v_t$ and 
$I(v_{t+1}; v_{t-1} \mid v_t) = 0$; a significant positive value rejects this 
null and shows that $v_{t-1}$ carries information about $v_{t+1}$ beyond 
$v_t$. This does not imply genuine higher-order memory in the underlying 
system: such signal commonly arises when a latent-state process is projected 
onto a lower-dimensional observed variable. We therefore treat 
$I(v_{t+1}; v_{t-1} \mid v_t) > 0$ as evidence that the observed state alone 
is insufficient, motivating the latent-variable model.

The plug-in estimator is positively biased at finite sample size, so we 
report the debiased value $\widehat{I}_{\mathrm{deb}} = \widehat{I} - 
\mathbb{E}_{\mathcal{H}_0}[\widehat{I}]$, with the null expectation obtained 
by a stratified permutation test ($B=500$): $v_{t-1}$ is shuffled within 
strata of $v_t$, preserving $P(v_t)$ and $P(v_{t+1}\mid v_t)$ while 
destroying dependence on $v_{t-1}$. Significance is assessed via 
$p = (1 + \sum_b \mathbf{1}[\widehat{I}^{(b)} \geq \widehat{I}_{\mathrm{obs}}])/(B+1)$. 
The test is applied separately within each formation, so a significant 
result reflects non-Markovianity intrinsic to that tactical context rather 
than an artifact of pooling heterogeneous formations.

\subsection{The Hyperedge-Latent Hidden Markov Model}
\label{sec:methods_hmm}

As formalized in Definitions~\ref{def:hmm_matrices}-\ref{def:hmm_matrices_B}, the model is fully
specified by three row-stochastic matrices: the assignment matrix
$S \in \mathbb{R}^{|V| \times K}$, with $S_{v,e}$ defining the probability
of selecting hidden state $e$ given the current vertex $v$; the emission
matrix $B \in \mathbb{R}^{K \times |V|}$, with $B_{e,v}$ representing the
probability that hidden state $e$ emits vertex $v$; and the transition
matrix $T \in \mathbb{R}^{K \times K}$, with $T_{e,e'}$ representing a
vertex-independent affinity between hidden states $e$ and $e'$
(row-stochastic in isolation). Crucially, $T$ does not by itself give the
process's realized transition probability: as shown in
Eq.~\eqref{eq:phi_kernel}, $T$ combines multiplicatively with the
vertex-dependent factor $S_{v_t,e_{t+1}}$, with the result renormalized to
yield $\varphi$, the actual transition kernel governing
$e_t \to e_{t+1}$. This vertex-independence of $T$ is nonetheless what
makes it the primary carrier of hidden memory in the process, distinct
from the vertex-dependent reweighting applied through $\varphi$.

Each of the $K$ hidden states is identified with one hyperedge: we write
$e_t \in E$ for the hidden state occupied at
time $t$, so that $e_t = e$ means the process is, at time $t$, in the
hyperedge $e$. Under this identification, the matrices above
translate directly into hyperedge-level quantities: $S_{v,e}$ is the
probability that vertex $v$ activates hyperedge $e$, $B_{e,v}$ is the
probability that hyperedge $e$ emits vertex $v$, and $T_{e,e'}$ is the
vertex-independent affinity from hyperedge $e$ to hyperedge $e'$.

Together, these define the joint process $(v_t, e_t)_{t \geq 0}$ as a
first-order Markov chain on ${V} \times {E}$:
\begin{equation}
    P(e_{t+1}, v_{t+1} \mid v_t, e_{t})
    = \varphi(v_t, e_{t+1}, e_{t}) \cdot B_{e_{t+1}, v_{t+1}},
    \label{eq:joint_kernel}
\end{equation}
where the hyperedge is selected according to
\begin{equation}
    \varphi(v_t, e_{t+1}, e_{t})
    = \frac{T_{e_{t}, e_{t+1}} \, S_{v_t, e_{t+1}}}
           {\sum_{\tilde{e} \in \mathcal{E}} T_{e_{t}, \tilde{e}} \, S_{v_t, \tilde{e}}}.
    \label{eq:phi_kernel}
\end{equation}

Although the joint chain is first-order Markov, the observed vertex
trajectory alone need not be: since $S_{v_t,\cdot}$ generally assigns
nonzero probability to several hyperedges, the current vertex $v_t$ does
not uniquely determine the hidden state $e_t$, so earlier vertices in the
trajectory can carry information about $e_t$ beyond what $v_t$ alone
reveals. Through $\varphi$'s dependence on $e_t$ (Eq.~\eqref{eq:phi_kernel}),
this extra information about past hyperedge occupancy can in turn
influence the distribution of $v_{t+1}$, making the marginal process
$(v_t)_{t\geq0}$ generally non-Markovian even though $(v_t,e_t)_{t\geq0}$
is. This residual memory is thus
governed entirely by $T$. When $T$ is uniform, $e_t$ carries no information 
about $e_{t-1}$ and the observed process reduces to first order (see the SB 
null model below); as $T$ departs from uniformity, the hyperedge sequence 
retains memory not recoverable from $v_t$ alone. Higher-order interactions can thus produce non-Markovianity, but not every 
non-Markovian process is higher-order, which is why the test is followed by a 
model comparison (Methods~\ref{sec:methods_comparison}).

\subsection{Parameters Estimation and Optimization}
\label{sec:methods_fitting}

The parameters of all models are estimated by maximizing the marginal log-likelihood of the observed sequence $V_{1:T}$, $P(v_2,\dots,v_T \mid \theta ,v_1)$ with $\theta$ the model parameters (e.g., $T,S,B$ for HMM). Depending on whether the model relies on a fully observed state space or unobserved latent variables, optimization employs either direct empirical estimation or the Expectation-Maximization (EM) framework.

\subsubsection{Direct Estimation for Standard Markov Chains (MC1, MC2)}
The standard first-order (MC1) and second-order (MC2) Markov chains lack latent hyperedge structure. Consequently, their transition probabilities can be estimated directly in closed form using empirical frequencies. 

To prevent zero-probability assignments for unobserved transitions, we apply Laplace smoothing:
\begin{equation}
  \hat{P}_{c,c'} = \frac{n_{c,c'} + \alpha}{\sum_{c''} n_{c,c''} + \alpha |\mathcal{C}|},
\end{equation}
where $c$ is the origin state (the previous vertex for MC1, or the pair of previous vertices for MC2), $c'$ is the destination vertex, $n_{c,c'}$ is the observed transition count, and $\alpha$ is the smoothing coefficient.
We use $\alpha = 5 \times 10^{-1}$, and we have checked that
other values ($5 \times 10^{-2}$, $5 \times 10^{-4}$) 
give consistent results. 

\subsubsection{Expectation-Maximization for the HMM and SB Null}
Because the sequence of traversed hyperedges is unobserved in both the full HMM and the memoryless SB null model, we estimate their parameters utilizing a generalized Expectation-Maximization (EM) algorithm.

\paragraph{Forward-Backward Inference (E-Step).}
Evaluating the likelihood and computing the posterior distribution over latent states is achieved via the forward-backward algorithm operating in $\mathcal{O}(T K^2)$ time. We define the forward variable $\alpha_t(e) \equiv \Prb(v_2,\dots,v_t,\, e_t=e\mid \theta, v_1)$ and the backward variable $\beta_t(e) \equiv \Prb(v_{t+1},\dots,v_T \mid e_t=e, \theta)$, with $\theta$ representing the parameters of the model.

With the initialization $\alpha_1(e) = S_{v_1, e}$, the forward recursion is given by:
\begin{equation}
\alpha_{t+1}(e) = B_{e,\,v_{t+1}} \sum_{e'} \alpha_t(e')\,\varphi_{e'\to e}(v_t),\qquad t=1,\dots,T-1.
\label{eq:forward}
\end{equation}
With $\beta_T(e)=1$, the backward recursion is:
\begin{equation}
\beta_t(e) = \sum_{e'} \varphi_{e\to e'}(v_t)\, B_{e',v_{t+1}}\,\beta_{t+1}(e').
\label{eq:backward}
\end{equation}
To avoid numerical underflow, $\alpha_t$ is rescaled at each step by $c_t = \sum_e \alpha_t(e) = \Prb(v_t\mid v_{1:t-1})$, yielding the stabilized, per-step normalized log-likelihood $\mathcal{L}_{\mathrm{HMM}} = \frac{1}{T-1}\sum_{t=2}^{T} \log c_t$. The posterior responsibilities $\gamma_t(e)$ and $\xi_t(e',e)$ are computed standardly from the normalized products of $\alpha$ and $\beta$.

\paragraph{The $Q$ Function.}
The conditional-data log-likelihood decomposes as a sum of local
terms, this is the key computational advantage of working with $Q$
rather than $P(v_2,\dots,v_T \mid \theta ,v_1)$ directly:
\begin{equation}
\begin{aligned}
    \log P(v_{2:T}, E \mid v_1, \theta)
    &= \log \left( P(e_1 \mid v_1) \prod_{t=2}^{T} P(e_t \mid e_{t-1}, v_{t-1}) P(v_t \mid e_t) \right) \\
    &= \underbrace{\log S_{v_1, e_1}}_{\text{init}}
    + \sum_{t=2}^{T} \underbrace{\log \varphi_{e_{t-1} \to e_t}(v_{t-1})}_{\text{transition}}
    + \sum_{t=2}^{T} \underbrace{\log B_{e_t, v_t}}_{\text{emission}}
\end{aligned}
\label{eq:complete_ll}
\end{equation}
Each term involves the hidden state at only one or two consecutive
timesteps. Taking the expectation under the posterior
$P(E \mid O, \theta^{\mathrm{old}})$ replaces the unknown indicator
functions of $e_t$ with the posterior responsibilities $\gamma_t(e)$ and
$\xi_t(e',e)$ computed in the E-step:
\begin{equation}
    Q(\theta \mid \theta^{\mathrm{old}})
    = \sum_{e} \gamma_1(e) \log S_{v_1, e}
    + \sum_{t=2}^{T} \sum_{e', e} \xi_{t-1}(e', e) \log \varphi_{e' \to e}(v_{t-1})
    + \sum_{t=2}^{T} \sum_{e} \gamma_t(e) \log B_{e, v_t}
    \label{eq:Q_function}
\end{equation}
Substituting the expression for $\varphi$ from Eq.~\eqref{eq:phi_kernel}
gives the fully expanded objective maximized in the M-step:
\begin{equation}
    Q(\theta \mid \theta^{\mathrm{old}})
    = \underbrace{\sum_{e} \gamma_1(e) \log S_{v_1,e}}_{\text{init}}
    + \underbrace{\sum_{t=2}^{T} \sum_{e', e} \xi_{t-1}(e', e) \log
    \frac{T_{e',e} \cdot S_{v_{t-1},e}}{\displaystyle\sum_{\tilde{e}} T_{e',\tilde{e}} \cdot S_{v_{t-1},\tilde{e}}}}_{\text{transition}}
    + \underbrace{\sum_{t=2}^{T} \sum_{e} \gamma_t(e) \log B_{e,v_t}}_{\text{emission}}
    \label{eq:Q_function_expanded}
\end{equation}
subject to the row-stochasticity constraints $\sum_e S_{v,e} = \sum_v B_{e,v} = 1$.

\paragraph{Generalized Baum-Welch Updates (M-Step).}
The M-step maximizes $Q$ (Eq.~\eqref{eq:Q_function_expanded}), subject to simplex constraints. The emission matrix $B$ admits a closed-form update in both the HMM and the SB null:
\begin{equation}
B_{e,v}^{(\text{new})} = \frac{\sum_{t=1}^{T-1}\gamma_{t+1}(e)\,\mathbb{1}[v_{t+1}=v]}{\sum_{t=1}^{T-1}\gamma_{t+1}(e)}.
\label{eq:Mstep_B}
\end{equation}
When the transition matrix $T$ is constrained to be row-uniform (the SB null model), the assignment matrix $S$ similarly admits a guaranteed, closed-form optimal update.

In the full HMM, $T$ and $S$ are coupled through the non-linear denominator of the transition kernel $\varphi$, precluding closed-form updates. We optimize them jointly via projected gradient ascent in logit space using the Adam optimizer (learning rate fixed at $0.05$ for both matrices, $\beta_1=0.9$, $\beta_2=0.999$, with a maximum of $10,000$ iterations). Let $\xi^{\text{row}}_t(e')=\sum_e \xi_t(e',e)$ and $\xi^{\text{col}}_t(\hat e)=\sum_{e'}\xi_t(e',\hat e)$. The gradients of $Q$ are:
\begin{align}
\frac{\partial Q}{\partial T_{e',\hat e}} &= \frac{\sum_t \xi_t(e',\hat e)}{T_{e',\hat e}} - \sum_t \frac{\xi^{\text{row}}_t(e')\,S_{v_t,\hat e}}{\sum_{\tilde e} T_{e',\tilde e}\,S_{v_t,\tilde e}}, \label{eq:gradT} \\[2pt]
\frac{\partial Q}{\partial S_{v,\hat e}} &= \mathbb{1}[v_1=v]\,\frac{\gamma_1(\hat e)}{S_{v,\hat e}} + \sum_{t=2}^{T}\mathbb{1}[v_{t-1}=v]\!\left[\frac{\xi^{\text{col}}_{t-1}(\hat e)}{S_{v,\hat e}} - \sum_{e'}\frac{T_{e',\hat e}\,\xi^{\text{row}}_{t-1}(e')}{\sum_{\tilde e} T_{e',\tilde e}\,S_{v,\tilde e}}\right]. \label{eq:gradS}
\end{align}
Row-stochasticity is maintained via softmax projection, with matrix values clipped at $10^{-12}$ to ensure strict numerical stability during gradient scaling.

\paragraph{Initialization and Symmetry Breaking.}
To assess convergence and stability under the least favorable initialization, the full HMM adopts a fully random initialization strategy rather than a warm start informed by the SB null model. The assignment and emission matrices ($S$ and $B$) are drawn entirely from uniform Dirichlet noise, discarding the converged SB parameters altogether; this constitutes the hardest possible starting condition, with no information from the null model carried forward. The transition matrix $T$ is likewise initialized uniformly. To break latent state symmetries, Gaussian noise ($\mathcal{N}(0, 0.8)$ for $T$ and $\mathcal{N}(0, 0.1)$ for $S$) is injected directly into the logit space prior to the first softmax projection.

\paragraph{Convergence Criterion.}
Both the SB null and the full HMM are trained with validation-based early
stopping. A held-out validation subset is evaluated every $5$ iterations;
training halts if the validation log-likelihood fails to improve by more
than $10^{-4}$ over $10$ consecutive evaluations (i.e.\ $50$ iterations
without improvement), and the parameters from the iteration with the best
observed validation log-likelihood are restored. These settings are identical for SB and HMM. When no
validation split is used (full-dataset fits), this early-stopping
mechanism is disabled and training instead runs for the full
$n_{\mathrm{iter}}$ iterations.

\subsection{The null models.}
\label{sec:methods_nulls}

We use two families of null models, each removing a different ingredient of 
the full HMM.

\paragraph{Removing latent memory (SB).} Setting $T$ to the uniform matrix,
$T_{e,e'} = 1/|E|$ for all pairs, removes the latent temporal memory: the
hyperedge selection in Eq.~\eqref{eq:phi_kernel} collapses to
$\varphi(v_t, e_t) = S_{v_t, e_t}$, the hyperedge is redrawn independently at
every step conditioned only on the current vertex, and the observed process
reduces to a first-order Markov chain on ${V}$ with transition matrix
$P_{\mathrm{SB}} = S B$. Restricting the support of $S$ and $B$ to a fixed,
known hyperedge incidence structure (i.e. $S_{v,e}=0$ for $v \notin e$ and
$B_{e,v}=0$ for $v \notin e$), together with the corresponding edge-weight
normalization, recovers the edge-dependent vertex-weighted (EDVW) random
walk on hypergraphs of Chitra and Raphael~\cite{chitra2019} as a special
case, where a walker samples an incident hyperedge proportionally to edge
weight and then a destination vertex within it; here, by contrast, $S$ and
$B$ are unconstrained and learned directly from data, with no incidence
structure imposed. Setting $K=|V|$ further collapses SB to an ordinary
first-order Markov chain, recovering MC1 as a special case. SB is thus the
strongest hypergraph-consistent baseline: any gain of the full HMM over SB
isolates the contribution of the latent higher-order structure $T$
specifically.

\paragraph{Removing hypergraph structure (MC1, MC2).} We also compare directly against standard Markov chains of order one (MC1) and order two (MC2). 
These baselines discard the latent hyperedge architecture entirely and model dynamics directly on the observed vertex set ${V}$. 
The MC1 model captures purely memoryless node-to-node transitions parameterized by a standard $|V| \times |V|$ stochastic matrix. 
The MC2 model incorporates a single step of historical dependence by extending the state space to pairs of successive vertices $(v_{t-1}, v_t)$, defining transitions to the next vertex $v_{t+1}$ via an expanded $|V|^2 \times |V|$ matrix.

\subsection{Model Comparison}
\label{sec:methods_comparison}

The hyperedge-latent HMM is benchmarked against the memoryless SB null model and the standard Markov chains (MC1, MC2) using out-of-sample predictive log-likelihood and in-sample complexity penalization (BIC).

\paragraph{Out-of-sample likelihood comparison.}
Out-of-sample log-likelihood is evaluated using a $10$-fold cross-validation scheme partitioned at the trajectory level, with a validation split extracted within each fold for early stopping, applied identically to the SB null and the full HMM as described in Section~\ref{sec:methods_fitting} (Convergence Criterion). Following termination, the parameter state maximizing the validation likelihood is restored for final test set evaluation.

Identical fold partitions are maintained across all models to ensure strictly paired comparisons. An empirical evaluation of model stability across multiple independent random seeds is provided in~\ref{sec:appendix_mega_plot_seeds_logos}.

Statistical significance of the paired likelihood differences is assessed using the Nadeau--Bengio corrected resampled $t$-test \cite{nadeau_inference_2003}. This correction inflates the variance estimate to account for statistical dependence stemming from overlapping training sets.

\paragraph{Complexity penalization and model selection (BIC).}
The number of latent hyperedges $K$ is determined using the Bayesian Information Criterion: $\mathrm{BIC} = -2\,\mathcal{L} + k \log n$, where $k$ is the number of free parameters and $n$ is the total number of observed transitions. Unlike the out-of-sample comparison above, BIC is not computed under cross-validation: each model is fit exactly once, on the entire dataset with no held-out split and no folding, and $\mathcal{L}$, $k$, and $n$ are likewise evaluated on that single full-dataset fit. Early stopping is correspondingly not applied when computing BIC; each model is instead trained for a fixed number of iterations.

Parameter counts are defined by the stochastic constraints of each model: $k_{\mathrm{SB}} = |V|(K-1) + K(|V|-1)$ for the memoryless null model, $k_{\mathrm{HMM}} = |V|(K-1) + K(|V|-1) + K(K-1)$ for the full HMM, $k_{\mathrm{MC1}} = |V|(|V|-1)$ for the first-order chain, and $k_{\mathrm{MC2}} = |V|^2(|V|-1)$ for the second-order chain. Because MC2 conditions on a pair of preceding vertices, its transition count $n$ is computed over one fewer step per trajectory than the other three model families. BIC serves strictly as an in-sample penalization metric for model selection. 
Formal statistical inference regarding model superiority is restricted to the out-of-sample likelihood evaluations.

\subsection{Synthetic Hypergraphs and Trajectory Generation}
\label{sec:methods_synthetic}

Synthetic trajectories are generated directly from the generative kernel defined in Eqs.~\eqref{eq:joint_kernel}--\eqref{eq:phi_kernel} to establish an exact parametric ground truth $(S, B, T)$. 
A random hypergraph is initialized with edge-dependent vertex weights (EDVW) \cite{chitra_random_2019} encoded via two incidence matrices.
The weight matrix $W \in \mathbb{R}^{|V| \times |E|}$ is strictly defined such that $W_{v,e}$ equals the size of hyperedge $e$ if $v \in e$, and $0$ otherwise. 
The emission weights matrix $R \in \mathbb{R}^{|E| \times |V|}$ can be constructed under one of two configurations, identical in every other respect. In the default configuration presented here, $R_{e,v}$ is drawn for each $v \in e$ from a truncated normal distribution ($\mu=1$, $\sigma=0.25$, bounded at $>0$) and $R_{e,v}=0$ otherwise, yielding a heterogeneous emission distribution within each hyperedge. Results under the alternative configuration, in which $R_{e,v}=1$ for each $v \in e$, yielding a uniform emission distribution within each hyperedge, are reported in \ref{sec:Supplementary-homogeneous}.
The ground-truth assignment and emission matrices are derived via row-normalization: $S_{v,e} = W_{v,e} / \sum_{e'} W_{v,e'}$ and $B_{e,v} = R_{e,v} / \sum_{v'} R_{e,v'}$; under the default configuration, this gives $B_{e,v} = 1/|e|$ for every member vertex of hyperedge $e$. This parameterization strictly scales $S$ by hyperedge size, while $B$ is heterogeneous (or uniform, under the alternative configuration) over each hyperedge's members. 
Note that, by construction, $S$ and $B$ share the same support under either configuration; correspondingly, the Jaccard similarity between their binarized versions is 1.

The ground-truth affinity matrix $T$ is defined as a persistence mixture:
\begin{equation}
    T = p_{\mathrm{stay}} \, I_E + (1 - p_{\mathrm{stay}}) \, \tfrac{1}{E} \mathbf{1}\mathbf{1}^\top.
\end{equation}
The parameter $p_{\mathrm{stay}} \in [0,1]$ modulates hyperedge persistence, and hence the magnitude of the induced non-Markovianity: $p_{\mathrm{stay}}=0$ yields the strictly memoryless SB process, while larger values produce a stronger higher-order signal. Because the HMM likelihood has no closed-form maximiser and is optimised iteratively by gradient-based updates (Methods~\ref{sec:methods_fitting}), the weaker this signal, the more trajectories are needed for the optimisation to recover the hyperedge structure reliably.

Sequence generation is initialized at a starting vertex $v_0$, with the initial hyperedge drawn as $e_0 \sim S_{v_0, \cdot}$. 
Trajectories are then extended for $\ell=30$ steps by iterating the generative kernel of Eqs.~\eqref{eq:joint_kernel}--\eqref{eq:phi_kernel}: $e_{t+1} \sim \varphi(v_t, \cdot, e_t)$ and $v_{t+1} \sim B_{e_{t+1}, \cdot}$, with $p_{\mathrm{stay}}=0.35$ (a second value, $p_{\mathrm{stay}}=0.20$, is tested in Supplementary Section~\ref{sec:appendix_homogeneous_power}). 
For each of the $V$ starting vertices, $N=200$ independent trajectories are sampled, yielding exactly $|V| \times N$ sequences of length $\ell+1$ per configuration. 
The likelihood of this data evaluated under the ground-truth parameters serves as the theoretical performance ceiling for all subsequent model benchmarking.

\subsection{Empirical Trajectory Extraction}
\label{sec:methods_empirical}

Empirical trajectories are extracted from professional football event data 
provided by StatsBomb. For each match, this dataset records a chronological 
sequence of timestamped on-ball events (passes, carries, shots, duels, 
pressures, etc.), each annotated with its type, the team and player involved, 
the player's tactical role at that moment, and its location on the pitch, 
together with the announced starting lineup and formation of each team. 
The pitch is discretised into a set of tactical roles $V$ (e.g., Right Center
Back, Left Defensive Midfield, Center Forward), with $|V|=24$ representing
the maximum number of roles observed; the effective vertex count for a
given team may be lower, since a role that never appears across the
extracted matches is excluded. A complete alphabetical enumeration of all included roles and their abbreviations is provided in \ref{sec:appendix_tactical_roles}.
Trajectories are then built from each team's uninterrupted possession sequences: 
each sequence is mapped to the successive roles of the players holding the ball, 
yielding a discrete trajectory over the role set $V$. These trajectories constitute 
the \textit{observed process}.

We followed the sames guidelines as in \cite{pena_markovian_nodate} for the maximum length of the trajectories, which is set to $\ell=30$ and the definition of a trajectory itself, which is defined as a sequence of uninterrupted possession events.
Because the number of roles exceeds the number of players on the pitch, a single player typically occupies 
several roles over the course of a match, and the discovered hyperedges are 
therefore to be interpreted as co-activation patterns of roles, that is, 
tactical units, rather than as fixed groups of players.

To illustrate the resulting datasets, Atl\'etico Madrid contributed 46
matches with formation 4-4-2, comprising a total of 5002 pass
trajectories with a mean length of $6.4 \pm 4.0$ steps across $|V|=21$
unique nodes; Bayern Munich contributed 7 matches with formations 3-4-2-1
and 3-5-1-1, comprising 713 trajectories with a mean length of
$8.4 \pm 5.8$ steps across $|V|=23$ unique nodes. Information about others datasets teams and formations is available in~\ref{sec:appendix_mega_plot_formations_logos}.

Prior to any model fitting, every formation is subjected to the 
conditional mutual information test described in 
Section~\ref{sec:methods_cmi}, which certifies within-formation 
non-Markovianity and rules out the possibility that pooling formations creates 
a spurious mixture-of-Markov-chains signal.

\section*{Data and code availability}
All code is available at \url{https://github.com/Arthurondeau/HyperDynamics}. Football trajectories are derived from the StatsBomb open dataset.

\section*{Acknowledgements}
We thank the StatsBomb team for making their football event data publicly available and Professor Bastien Chopard for fruitful discussions.
\bibliography{references}

\clearpage
\section*{Supplementary Material}

\setcounter{section}{0}
\setcounter{table}{0}
\setcounter{equation}{0}
\setcounter{figure}{0}
\renewcommand{\figurename}{Supplementary Figure}
\renewcommand{\tablename}{Supplementary Table}
\renewcommand{\thesection}{Supplementary Note \arabic{section}}

\section{Synthetic homogeneous Watts–Strogatz hypergraph}
\label{sec:Supplementary-homogeneous}
Here we present the results in Supplementary Figure~\ref{fig:synth-homo-reports-and-params} 
and Supplementary Table~\ref{tab:model_comparison_k15_p_0.35} for the synthetic 
Watts--Strogatz hypergraph configuration in which $R$ is constructed 
differently from the main text. In the main text, $R_{e,v}$ is drawn from a 
truncated normal distribution ($\mu=1$, $\sigma=0.25$, bounded at $>0$) for 
each $v \in e$. Here, instead, $R_{e,v}=1$ for each $v \in e$ and $0$ 
otherwise, yielding a uniform emission distribution within each hyperedge. The 
learned emission matrix $\hat{B}_{e,v}$ reproduces the block structure 
of the ground-truth incidence matrix (red boxes), with each recovered 
hyperedge concentrating its probability mass on the correct set of member 
vertices (Supplementary Fig.~\ref{fig:synth-homo-reports-and-params}c), with one exception: hyperedge 10.
This exception directly illustrates a consequence of optimizing the HMM with gradient descent: convergence to the true solution at $K=15$ is not guaranteed.

Thus, to assess whether the hyperedges recovered at the BIC-optimal $K$ remain 
stable as $K$ increases (even in a non-optimal case where not all the ground truth hyperedges are recovered at the optimal $K=15$), we examine two distinct questions. First, does a 
larger $K$ \emph{refine} the hyperedges found at a smaller $K$, splitting 
them into coherent sub-groups, or does it instead produce unrelated 
structure? Second, independently of any comparison across $K$, are the 
hyperedges recovered \emph{at a given} $K$ mutually distinct, or does 
increasing $K$ simply duplicate existing groups? We report these two 
diagnostics separately in Supplementary Tables~\ref{tab:cross_K_refinement} and 
\ref{tab:within_K_redundancy}, since they answer different questions and 
are not directly comparable.

\paragraph{Cross-$K$ refinement.} Supplementary Table~\ref{tab:cross_K_refinement} 
compares each $K_{\mathrm{large}}$ with the 
smaller $K_{\mathrm{small}}$ that precedes immediately in the sequence. \emph{Subset} is the fraction 
of non-empty hyperedges at $K_{\mathrm{large}}$ whose support is entirely 
contained within the support of at least one hyperedge at 
$K_{\mathrm{small}}$, i.e.\ the fraction that are a genuine refinement of 
some coarser group rather than an unrelated recombination of roles. 
\emph{Unique} restricts this to hyperedges with exactly one such 
$K_{\mathrm{small}}$ parent, as opposed to several candidate parents, and 
measures how unambiguous this lineage is. \emph{Orphaned} is the fraction 
of $K_{\mathrm{small}}$ hyperedges that are not refined by \emph{any} 
hyperedge at $K_{\mathrm{large}}$, i.e.\ structures that disappear when $K$ 
is increased. \emph{Support} reports the mean number of roles per hyperedge 
before and after the transition.

\paragraph{Within-$K$ redundancy.} Supplementary Table~\ref{tab:within_K_redundancy} 
instead asks, independently of any smaller $K$, whether hyperedges at a 
given $K$ overlap with one another. \emph{Avg.\ Jaccard} is the mean, over 
hyperedges at that $K$, of each hyperedge's best Jaccard overlap with any 
\emph{other} hyperedge at the same $K$; higher values indicate that 
hyperedges share roles rather than remaining distinct. \emph{Frac.\ dup.} 
is the fraction of hyperedges with at least one such near-duplicate partner 
(Jaccard $\geq 0.8$). \emph{Support} again reports the mean number of roles 
per hyperedge.

We observe that increasing $K$ causes the hyperedges correctly recovered at 
$K=15$ to split into smaller, overlapping subsets rather than into cleanly 
distinct refinements: the fraction of unique-parent matches degrades 
steadily (Supplementary Table~\ref{tab:cross_K_refinement}), while the fraction of 
near-duplicate hyperedges at a given $K$ rises in parallel 
(Supplementary Table~\ref{tab:within_K_redundancy}). In other words, beyond the 
BIC-optimal $K$, additional hyperedges are largely redundant copies of existing structures, as also shown in Supplementary Figure \ref{fig:B_23_ground_truth_match}.

\begin{table}[htbp]
  \centering
  \begin{threeparttable}
    \caption{Model comparison for homogeneous hypergraph with $p_{stay}=0.35$ at true $K$ ($K=15$) and at its largest
    BIC-dominant $K=23$. Test-set log-likelihood (mean $\pm$ 1 s.d.
    across $n=10$ cross-validation folds), BIC, and the difference in
    test log-likelihood relative to MC2 ($\alpha=0.5$), the strongest
    baseline.    \label{tab:model_comparison_k15_p_0.35}
} 
  \setlength{\tabcolsep}{27pt}
  \begin{tabular}{l c c c}
  \toprule
  \textbf{Model} & \textbf{Test LL (nats)} & \textbf{BIC} & {$\Delta$} \\
  \midrule
  Null Model (SB) & $-2.261 \pm 0.002$ & \num{410859} & -0.216 \\
  MC1 ($\alpha=0.5$) & $-2.253 \pm 0.002$ & \num{407910} & -0.208 \\
  MC2 ($\alpha=0.5$) & $-2.046 \pm 0.002$ & \num{403593} & ref. \\
  \textbf{HMM ($K=15$)} & $-2.027 \pm 0.002$ & \textbf{\num{368601}} & +0.019*** \\
  HMM ($K=23$)$^\dagger$ & $\mathbf{-2.006 \pm 0.001}$ & \num{374381} & +0.040*** \\
  \midrule
  \textit{Ground Truth} & $-2.005 \pm 0.001$ & \num{368380} & -- \\
  \bottomrule
  \end{tabular} 
  \begin{tablenotes}\footnotesize
  \item[$\dagger$] $K=23$ is the largest $K$ at which the HMM's BIC (computed on a single fit to the full dataset) is lower than or equal to every baseline's full-dataset BIC. This is a descriptive comparison of point estimates, not a significance claim.
  \item[*/**/***] $p<0.05/0.01/0.001$ (one-sided Nadeau--Bengio corrected paired $t$-test on test log-likelihood; $n=10$ folds).
  \item[$\Delta$] Difference in mean test log-likelihood relative to MC2 ($\alpha=0.5$), the strongest baseline (positive favors the row's model). This is the only valid basis for a significance claim, since it is computed on paired, fold-matched differences; the mean $\pm$ 1 s.d. reported per row describes each model's own fold-to-fold variability and should not be compared for overlap. BIC values are point estimates from a single full-dataset fit and are not subject to this test.
  \item[] Ground truth is shown as an oracle reference (upper bound on achievable performance) and is not subject to a significance test.
  \end{tablenotes}
  \end{threeparttable}
  \end{table}
  
\begin{table}[h!]
  \centering
  \setlength{\tabcolsep}{19pt}
  \begin{tabular}{c c c c c}
  \toprule
  $K_{\mathrm{small}} \to K_{\mathrm{large}}$ & Subset & Unique & Orphaned & Support ($e \to e'$) \\
  \midrule
  $15 \to 17$ & 100.0\% & 94.1\% & 6.7\%  & 4.1 $\to$ 3.9 \\
  $17 \to 19$ & 78.9\%  & 73.3\% & 5.9\%  & 3.9 $\to$ 3.8 \\
  $19 \to 21$ & 95.2\%  & 85.0\% & 15.8\% & 3.8 $\to$ 4.0 \\
  $21 \to 23$ & 100.0\% & 69.6\% & 4.8\%  & 4.0 $\to$ 3.8 \\
  \bottomrule
  \end{tabular}
  \caption{Cross-$K$ refinement: does each $K_{\mathrm{large}}$ refine the 
  hyperedges recovered at the immediately preceding $K_{\mathrm{small}}$? 
  See text for metric definitions.   \label{tab:cross_K_refinement}
}
\end{table}

\begin{table}[h!]
  \centering
  \setlength{\tabcolsep}{29pt}
  \begin{tabular}{c c c c}
  \toprule
  $K_{\mathrm{large}}$ & Avg. Jaccard & Frac. dup. & Support \\
  \midrule
  17 & 0.543 & 23.5\% & 3.94 \\
  19 & 0.470 & 21.1\% & 3.84 \\
  21 & 0.646 & 38.1\% & 4.05 \\
  23 & 0.688 & 43.5\% & 3.78 \\
  \bottomrule
  \end{tabular}
  \caption{Within-$K$ redundancy: are the hyperedges recovered at a given 
  $K$ mutually distinct, or do they duplicate one another (near-duplicate 
  threshold: Jaccard $\geq 0.8$)? See text for metric definitions.}
  \label{tab:within_K_redundancy}
\end{table}

\begin{figure}[h!]
  \centering
  \includegraphics[width=\textwidth]{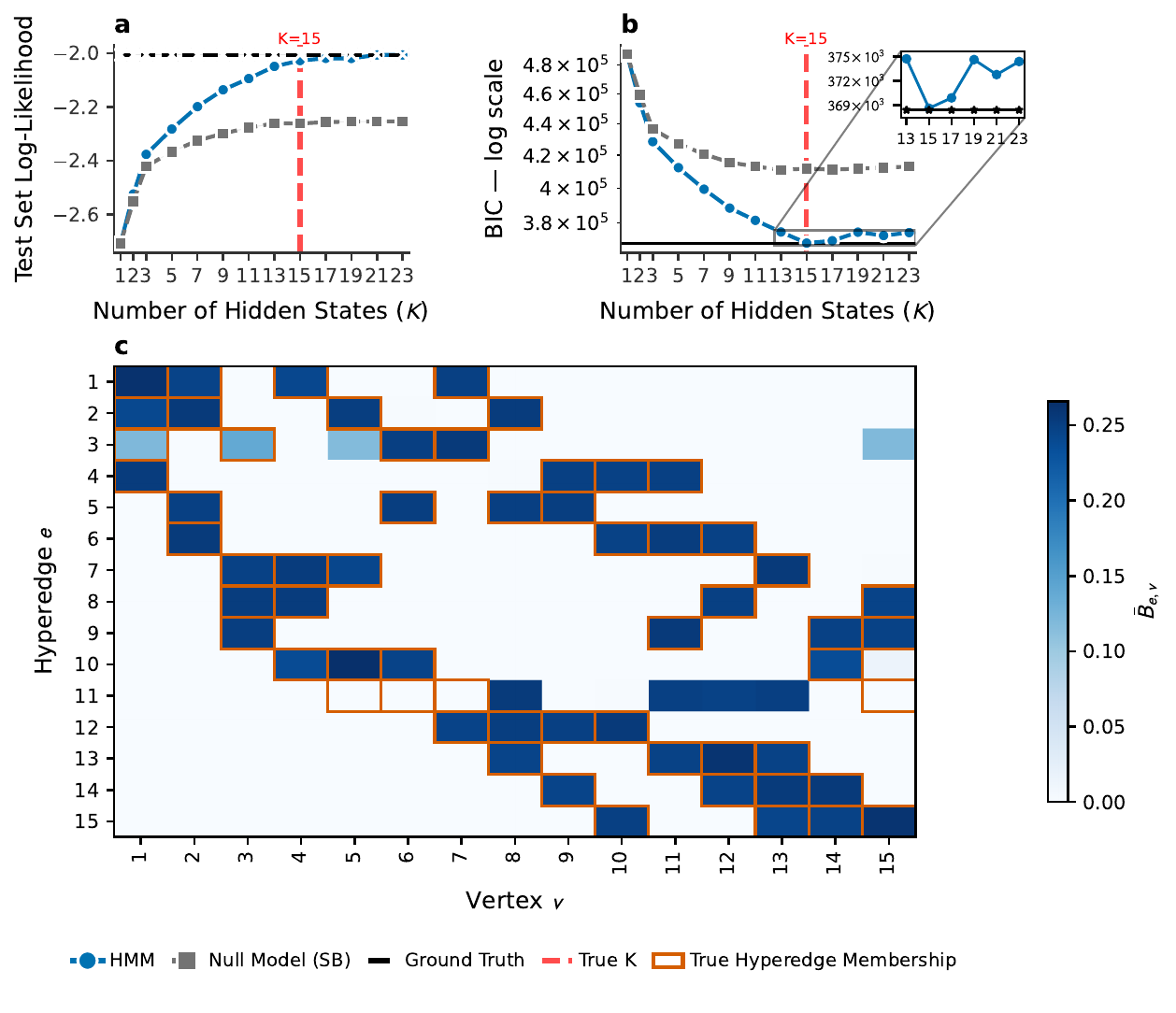}
\caption{\textbf{Model performance and recovered parameters on synthetic data}
  ($|E|=15$ hyperedges of size 4, $|V|=15$ vertices, Watts–Strogatz rewiring
  probability $p=10^{-3}$, and overlap parameter $l=2$ controlling the degree
  of vertex sharing between neighbouring hyperedges).
  \textbf{a}, Test-set log-likelihood per step as a function of the number of
  hidden states $K$, comparing the HMM against the SB null model and ground
  truth. Shaded bands show $\pm 1$ standard deviation across 10
  cross-validation folds. The dashed red line marks the true number of
  hidden states.
  \textbf{b}, Bayesian Information Criterion (BIC, log scale) across models
  as a function of $K$; lower is better.
  \textbf{c}, Fold-averaged emission matrix $\bar{B}_{e,v}$ at the ground-truth 
  $K=|E|$; rows are hierarchically clustered (Ward linkage) and aligned to the
  ground-truth incidence matrix; red boxes mark the ground-truth hyperedge
  support.    \label{fig:synth-homo-reports-and-params}
}
\end{figure}

\newpage

\begin{figure}[h]
    \centering
    \includegraphics[width=0.85\textwidth]{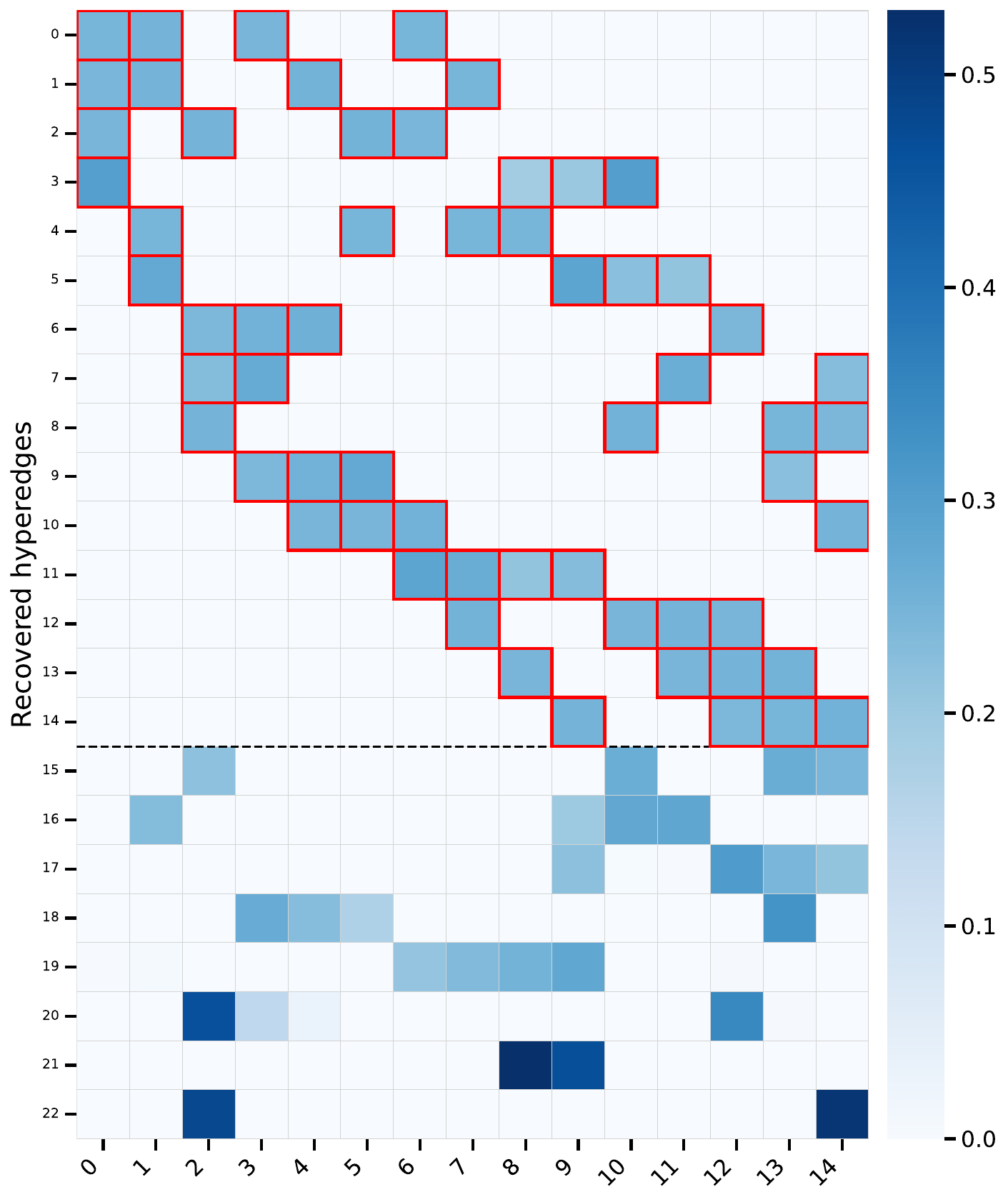}
    \caption{Correspondence between recovered states and ground-truth 
    hyperedges. The $23 \times 15$ emission matrix $\hat{B}$, with rows 
    reordered so that the first 15 states are matched one-to-one to the 
    15 ground-truth hyperedges (Hungarian assignment on Jensen--Shannon 
    similarity between each recovered state's emission distribution and 
    each ground-truth group's node membership); row labels give the 
    original state index. The remaining 8 states, below the dashed red 
    line, are unmatched and ordered by decreasing similarity to their 
    closest ground-truth group. Red boxes mark the ground-truth 
    hyperedge support for each of the 15 matched states.}
    \label{fig:B_23_ground_truth_match}
\end{figure}

\begin{figure*}[t]
  \centering
  \includegraphics[width=\textwidth]{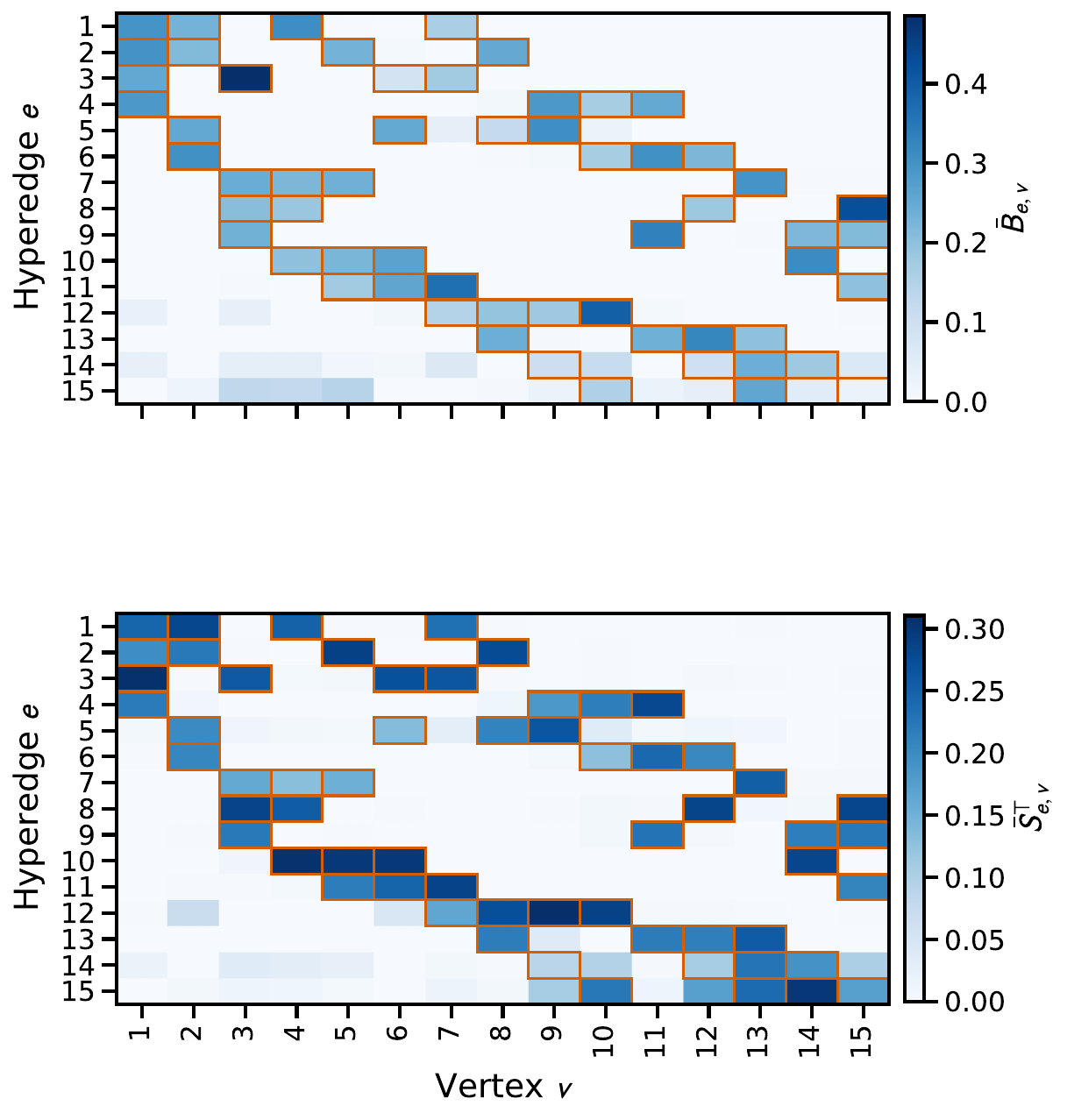}
  \caption{\textbf{Recovered emission and assignment structure on synthetic data}
  ($|E|=15$ hyperedges of size 4, $|V|=15$ vertices, Watts–Strogatz rewiring
  probability $p=10^{-3}$, and overlap parameter $l=2$ controlling the degree
  of vertex sharing between neighbouring hyperedges).
  \textbf{Top}, fold-averaged emission matrix $\bar{B}_{e,v}$; rows and
  columns are hierarchically clustered (Ward linkage) on the ground-truth
  incidence matrix $R$. \textbf{Bottom}, fold-averaged assignment matrix
  $\bar{S}^\top_{e,v}$, sharing the same hyperedge (row) and vertex
  (column) ordering as the top panel for direct visual comparison. In both
  panels, vermillion boxes mark the ground-truth hyperedge support. The two
  panels show that the recovered emission and assignment structures agree
  with each other and with the ground-truth incidence pattern.}
  \label{fig:synth_homo_panel1_SI}
\end{figure*}

We also report in Supplementary Figure ~\ref{fig:synth_homo_panel1_SI} both $\hat{S}$ and $\hat{B}$ learned matrices from the example in the main text (where $R$ was drawn from normal distribution). The Jaccard similarity between their binarized (significant) supports is 0.984, indicating that the recovered emission and assignment structures agree almost perfectly on which vertex-hyperedge pairs are significant, which was the expected result since $S$ and $B$ matrices in the synthetic process share the same support.

\clearpage

\section{Synthetic heterogeneous Watts–Strogatz hypergraph}
\label{sec:appendix_synthetic_heterogeneous}

\begin{figure}[htbp]
  \centering
  \includegraphics[width=0.95\textwidth]{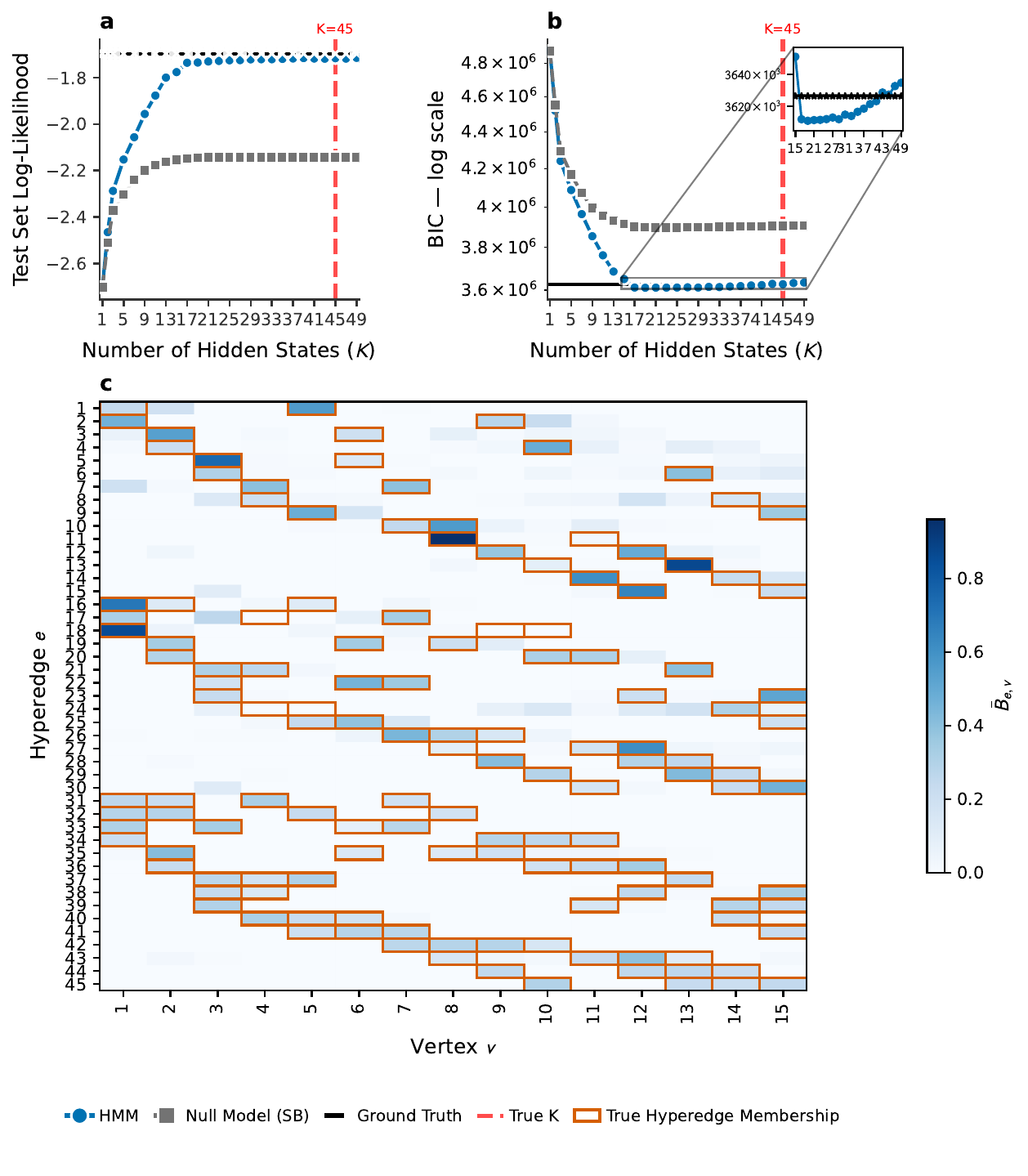}
  \caption{Model performance and recovered parameters on synthetic data
  ($p_{\mathrm{stay}} = 0.35$, $|E| = 45$ hyperedges of size $\{2,3,4\}$,
  $|V| = 15$ vertices, Watts--Strogatz rewiring probability $p = 10^{-3}$,
  and overlap parameter $l = 2$ controlling the degree of vertex sharing
  between neighbouring hyperedges)
  \textbf{a}, Test-set log-likelihood per step as a function of the number
  of hidden states $K$, comparing the HMM against the SB null model. The
  dashed red line marks the true number of hidden states.
  \textbf{b}, Bayesian Information Criterion (BIC, log scale) across models
  as a function of $K$; lower is better.
  \textbf{c}, Fold-averaged emission matrix $\bar{B}_{e,v}$ at the expected
  $K$, with vertices assigned to their most probable discovered tactical
  state; low-confidence assignments (Nadeau--Bengio corrected $t$-test) are
  marked as failed and boxed in dashed black. Ground truth membership is indicated with red boxes.}
  \label{fig:panel_k45}
\end{figure}

\begin{figure}[t]
    \centering
    \includegraphics[width=0.6\linewidth]{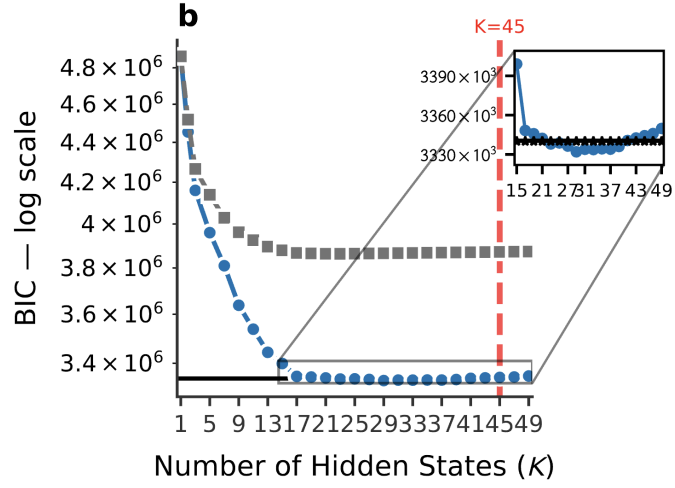}
    \caption{\textbf{BIC as a function of the number of hidden states for the heterogeneous hypergraph with ten times more samples.}
    BIC of the HMM (blue) and of the SB null model (grey) versus $K$, with the ground-truth reference (black) computed with its nominal number of parameters. The dashed red line marks the true $K=45$. The inset zooms on the region of the BIC minimum. With ten times more samples (2000 trajectories per node instead of 200), the minimum moves from $K\approx 17$--$21$ to $K\approx 29$--$33$, closer to the true number of hyperedges.}
    \label{fig:hetero_bic_2000samples}
\end{figure}

Here we present results for a heterogeneous hypergraph, with hyperedges of mixed sizes $\{2,3,4\}$ (Supplementary Figure~\ref{fig:panel_k45}, Supplementary Table~\ref{tab:model_comparison_k45}), using the same number of samples as in the homogeneous case.
The HMM at the true $K=45$ outperforms all baselines, including MC2 ($\Delta=+0.114$, $p<0.001$), and comes close to the ground truth.
The emission matrix recovers the ground-truth hyperedge membership.
In contrast to the homogeneous case, however, the BIC is minimised at a smaller number of hidden states ($K\approx 17$--$21$) than the true $K=45$.
This reflects a trade-off between model complexity and optimisation at a fixed sample size.
At $K=45$, the model has many more parameters to estimate, and strongly overlapping hyperedges induce nearly identical transition statistics, so that the gradient-based optimisation cannot resolve all of them from the available data.
Fewer states then reach a comparable likelihood at a lower parameter cost, which the BIC penalty favours.
Increasing the number of samples tenfold shifts the BIC minimum towards the true value, to $K\approx 29$--$33$ (Supplementary Figure~\ref{fig:hetero_bic_2000samples}), consistent with a larger fraction of the overlapping hyperedges becoming resolvable.
The recovered structure remains coherent with the ground truth in both cases, and the BIC of the HMM stays below that of all null models.
These results extend the conclusions of the main text beyond homogeneous hypergraphs to heterogeneous configurations.

\begin{table}[htbp]
\centering
\begin{threeparttable}
\caption{Model comparison at true $K$ ($K=45$) and at its largest
full-dataset BIC-dominant $K=49$. Test-set log-likelihood (mean $\pm$ 1 s.d.
across $n=10$ cross-validation folds), full-dataset BIC, and the difference in
test log-likelihood relative to MC2 ($\alpha=0.5$), the strongest
baseline.}
\label{tab:model_comparison_k45}
\setlength{\tabcolsep}{25pt}
\begin{tabular}{l c c c}
\toprule
\textbf{Model} & \textbf{Test LL (nats)} & \textbf{BIC} & {$\Delta$} \\
\midrule
MC1 ($\alpha=0.5$) & $-2.145 \pm 0.009$ & \num{3890553} & -0.231 \\
Null Model (SB) & $-2.144 \pm 0.009$ & \num{3896555} & -0.231 \\
MC2 ($\alpha=0.5$) & $-1.913 \pm 0.010$ & \num{3664997} & ref. \\
\textbf{HMM ($K=45$)} & $\mathbf{-1.846 \pm 0.011}$ & \textbf{\num{3627277}} & +0.067*** \\
HMM ($K=49$)$^\dagger$ & $-1.847 \pm 0.011$ & \num{3634724} & +0.067*** \\
\midrule
\textit{Ground Truth} & $-1.831 \pm 0.011$ & \num{3626343} & -- \\
\bottomrule
\end{tabular}
\begin{tablenotes}\footnotesize
\item[$\dagger$] $K=49$ is the largest $K$ at which the HMM's BIC (computed on a single fit to the full dataset) is lower than or equal to every baseline's full-dataset BIC. This is a descriptive comparison of point estimates, not a significance claim.
\item[*/**/***] $p<0.05/0.01/0.001$ (one-sided Nadeau--Bengio corrected paired $t$-test on test log-likelihood; $n=10$ folds).
\item[$\Delta$] Difference in mean test log-likelihood relative to MC2 ($\alpha=0.5$), the strongest baseline (positive favors the row's model). This is the only valid basis for a significance claim, since it is computed on paired, fold-matched differences; the mean $\pm$ 1 s.d. reported per row describes each model's own fold-to-fold variability and should not be compared for overlap. BIC values are point estimates from a single full-dataset fit and are not subject to this test.
\item[] Ground truth is shown as an oracle reference (upper bound on achievable performance) and is not subject to a significance test.
\end{tablenotes}
\end{threeparttable}
\end{table}

\clearpage

\section{Synthetic homogeneous Watts–Strogatz hypergraph Power Analysis}
\label{sec:appendix_homogeneous_power}
In this section we present the results for a lower probability to stay in the hyperedge $p_{stay}$ in the synthetic process, Supplementary Table \ref{tab:model_comparison_k15_p_0.2}.
We observe that for the a fixed number of samples used, when non Markovian signal magnitude (i.e $p_{stay}$) gets lower than the one observed in the main text, the HMM model's advantage is no longer observed at the expected $K$, here $K=15$. This result echoes the limitations of the framework, where, when the magnitude of the signal capturing the 
higher-order interactions is low, a sufficient number of samples are needed to train the HMM model.

\begin{table}[htbp]
  \centering
  \begin{threeparttable}
    \caption{Model comparison for homogeneous hypergraph with $p_{stay}=0.2$ at true $K$ ($K=15$) and at its largest
    BIC-dominant $K=23$. Test-set log-likelihood (mean $\pm$ 1 s.d.
    across $n=10$ cross-validation folds), BIC, and the difference in
    test log-likelihood relative to MC2 ($\alpha=0.5$), the strongest
    baseline.}
  \label{tab:model_comparison_k15_p_0.2}
  \setlength{\tabcolsep}{27pt}
  \begin{tabular}{l c c c}
    \toprule
    \textbf{Model} & \textbf{Test LL (nats)} & \textbf{BIC} & {$\Delta$} \\
    \midrule
    Null Model (SB) & $-2.261 \pm 0.003$ & \num{410151} & -0.112 \\
    MC1 ($\alpha=0.5$) & $-2.253 \pm 0.003$ & \num{407180} & -0.103 \\
    MC2 ($\alpha=0.5$) & $-2.149 \pm 0.003$ & \num{420260} & ref. \\
    HMM ($K=15$) & $-2.151 \pm 0.003$ & \textbf{\num{393140}} & -0.002 \\
    HMM ($K=23$)$^\dagger$ & $\mathbf{-2.137 \pm 0.003}$ & \num{396173} & +0.012*** \\
    \midrule
    \textit{Ground Truth} & $-2.137 \pm 0.003$ & \num{487609} & -- \\
    \bottomrule
    \end{tabular}
    \begin{tablenotes}\footnotesize
    \item[$\dagger$] $K=23$ is the largest $K$ at which the HMM's BIC (computed on a single fit to the full dataset) is lower than or equal to every baseline's full-dataset BIC. This is a descriptive comparison of point estimates, not a significance claim.
    \item[*/**/***] $p<0.05/0.01/0.001$ (one-sided Nadeau--Bengio corrected paired $t$-test on test log-likelihood; $n=10$ folds).
    \item[$\Delta$] Difference in mean test log-likelihood relative to MC2 ($\alpha=0.5$), the strongest baseline (positive favors the row's model). This is the only valid basis for a significance claim, since it is computed on paired, fold-matched differences; the mean $\pm$ 1 s.d. reported per row describes each model's own fold-to-fold variability and should not be compared for overlap. BIC values are point estimates from a single full-dataset fit and are not subject to this test.
    \item[] Ground truth is shown as an oracle reference (upper bound on achievable performance) and is not subject to a significance test.
    \end{tablenotes}
    \end{threeparttable}
    \end{table}

\clearpage

\section{Overlap and Hyperedge Size Influence on HMM Performance}
\label{sec:appendix_synthetic_overlap_size}

We study how the HMM advantage over the null model depends on two parameters of the Watts--Strogatz hypergraph generator: the overlap parameter $l$ between consecutive hyperedges (Supplementary Table~\ref{tab:overlap_sweep_k15}) and the hyperedge size $d$ (Supplementary Table~\ref{tab:hyperedge_size_sweep_k15}).
The advantage $\Delta$ increases monotonically with $l$.
It peaks at $d=3$ and decreases for larger hyperedges.
At $d=2$ hyperedges reduce to pairwise edges and the advantage vanishes as expected.

\begin{table}[htbp]
  \centering
  \begin{threeparttable}
  \caption{HMM advantage over the Null Model (SB) at $K=15$ across
  hyperedge overlap values. $\Delta$ is the mean paired difference in test
  log-likelihood (HMM $-$ SB), mean $\pm$ 1 s.d. across $n=10$
  cross-validation folds.}
  \label{tab:overlap_sweep_k15}
  \setlength{\tabcolsep}{70pt}
  \begin{tabular}{c c}
  \toprule
  \textbf{Hyperedge Overlap} & {$\Delta$} \\
  \midrule
  0 & $0.126 \pm 0.003$ \\
  1 & $0.165 \pm 0.002$ \\
  2 & $0.249 \pm 0.010$ \\
  3 & $0.261 \pm 0.004$ \\
  4 & $0.274 \pm 0.004$ \\
  5 & $\mathbf{0.279 \pm 0.005}$ \\
  \bottomrule
  \end{tabular}
  \begin{tablenotes}\footnotesize
  \item[] $\Delta$ increases monotonically with hyperedge overlap, with the
  largest HMM advantage (bolded) observed at overlap $=5$.
  \end{tablenotes}
  \end{threeparttable}
  \end{table}

  \begin{table}[htbp]
    \centering
    \begin{threeparttable}
    \caption{HMM advantage over the Null Model (SB) at $K=15$ across
    hyperedge sizes. $\Delta$ is the mean paired difference in test
    log-likelihood (HMM $-$ SB), mean $\pm$ 1 s.d. across $n=10$
    cross-validation folds.}
    \label{tab:hyperedge_size_sweep_k15}
    \setlength{\tabcolsep}{70pt}
    \begin{tabular}{c c}
    \toprule
    \textbf{Hyperedge Size} & {$\Delta$} \\
    \midrule
    2  & $0.000 \pm 0.000$ \\
    3  & $\mathbf{0.281 \pm 0.006}$ \\
    4  & $0.251 \pm 0.017$ \\
    6  & $0.159 \pm 0.010$ \\
    10 & $0.050 \pm 0.004$ \\
    \bottomrule
    \end{tabular}
    \begin{tablenotes}\footnotesize
    \item[] $\Delta$ peaks at hyperedge size $=3$ (bolded) and decreases for
    both smaller and larger hyperedges, with near-zero advantage at size
    $=2$.
    \end{tablenotes}
    \end{threeparttable}
    \end{table}

\newpage

\section{Negative Samples}
\label{sec:appendix_negative_synth}
First, we present the results on classic first order Markov process generated on a homogeneous Watts–Strogatz hypergraph, Supplementary Table \ref{tab:model_comparison_k15_p_0}.

  \begin{table}[htbp]
    \centering
    \begin{threeparttable}
    \caption{Model comparison for homogeneous hypergraph with $p_{stay}=0$ (Markovian process) at true $K$ ($K=15$) and at its largest
    BIC-dominant $K=15$. Test-set log-likelihood (mean $\pm$ 1 s.d.
    across $n=10$ cross-validation folds), full-dataset BIC, and the difference in
    test log-likelihood relative to MC1 ($\alpha=0.5$), the strongest
    baseline.}
    \label{tab:model_comparison_k15_p_0}
    \setlength{\tabcolsep}{27pt}
    \begin{tabular}{l c c c}
      \toprule
      \textbf{Model} & \textbf{Test LL (nats)} & \textbf{BIC} & {$\Delta$} \\
      \midrule
      Null Model (SB) & $-2.258 \pm 0.002$ & \num{410428} & -0.005 \\
      MC2 ($\alpha=0.5$) & $-2.254 \pm 0.001$ & \num{439946} & -0.001 \\
      \textbf{MC1} ($\alpha=0.5$) & $\mathbf{-2.253 \pm 0.001}$ & \textbf{\num{407378}} & ref. \\
      HMM ($K=15$) & $-2.256 \pm 0.001$ & \num{412735} & -0.003 \\
      \midrule
      \bottomrule
      \end{tabular}
      \begin{tablenotes}\footnotesize
      \item[$\dagger$] $K=15$ is the largest $K$ at which the HMM's BIC (computed on a single fit to the full dataset) is lower than or equal to every baseline's full-dataset BIC. This is a descriptive comparison of point estimates, not a significance claim.
      \item[$\Delta$] Difference in mean test log-likelihood relative to MC1 ($\alpha=0.5$), the strongest baseline (positive favors the row's model). This is the only valid basis for a significance claim, since it is computed on paired, fold-matched differences; the mean $\pm$ 1 s.d. reported per row describes each model's own fold-to-fold variability and should not be compared for overlap. BIC values are point estimates from a single full-dataset fit and are not subject to this test.
      \end{tablenotes}
      \end{threeparttable}
      \end{table}
We see that in Supplementary Table \ref{tab:model_comparison_k15_p_0}, MC1 correctly dominates: it attains the lowest BIC and the 
highest predictive likelihood, exactly as expected when the true generative 
process is first-order Markov process.

We also present in Supplementary Table \ref{tab:model_comparison_k15_mc2} 
the case where the synthetic 
process is genuinely generated by a second-order temporal mechanism. In that 
controlled setting, MC2 correctly dominates: it attains the lowest BIC and the 
highest predictive likelihood, exactly as expected when the true generative 
process is second-order temporal. The empirical football data shows the opposite pattern: MC2 is heavily penalised while the edge-latent HMM retains its advantage. This confirms that the higher-order dependencies in the real data stem from the latent hyperedge footprint rather than from second-order positional memory, since a second-order temporal process would have produced the MC2-dominated signature of the negative control.

\begin{table}[htbp]
  \centering
  \begin{threeparttable}
  \caption{Model comparison for homogeneous hypergraph with second order markov process at true $K$ ($K=15$) and at its largest
  BIC-dominant $K=15$. Test-set log-likelihood (mean $\pm$ 1 s.d.
  across $n=10$ cross-validation folds), BIC, and the difference in
  test log-likelihood relative to MC2 ($\alpha=0.5$), the strongest
  baseline.}
  \label{tab:model_comparison_k15_mc2}
  \setlength{\tabcolsep}{27pt}
  \begin{tabular}{l c c c}
    \toprule
    \textbf{Model} & \textbf{Test LL (nats)} & \textbf{BIC} & {$\Delta$} \\
    \midrule
    Null Model (SB) & $-2.266 \pm 0.004$ & \num{410050} & -0.552 \\
    MC1 ($\alpha=0.5$) & $-2.243 \pm 0.003$ & \num{405876} & -0.529 \\
    \textbf{MC2} ($\alpha=0.5$) & $\mathbf{-1.714 \pm 0.008}$ & \textbf{\num{345784}} & ref. \\
    HMM ($K=15$)& $-2.168 \pm 0.005$ & \num{396915} & -0.454 \\
    \midrule
    \bottomrule
    \end{tabular}
    \begin{tablenotes}\footnotesize
    \item[$\dagger$] $K=15$ is the largest $K$ at which the HMM's BIC (computed on a single fit to the full dataset) is lower than or equal to every baseline's full-dataset BIC. This is a descriptive comparison of point estimates, not a significance claim.
    \item[$\Delta$] Difference in mean test log-likelihood relative to MC2 ($\alpha=0.5$), the strongest baseline (positive favors the row's model). This is the only valid basis for a significance claim, since it is computed on paired, fold-matched differences; the mean $\pm$ 1 s.d. reported per row describes each model's own fold-to-fold variability and should not be compared for overlap. BIC values are point estimates from a single full-dataset fit and are not subject to this test.
    \end{tablenotes}
    \end{threeparttable}
    \end{table}
\newpage

\section{Markovianity Test For Empirical Data}
\label{sec:appendix_Markovianity_empirical}

All formations exhibit statistically significant non-Markovian structure 
($p < 0.01$ in all cases), confirming that the observed trajectories cannot 
solely be explained by a first-order Markov process. Crucially, this result holds 
\emph{within each individual formation} ruling out the alternative explanation 
that the apparent non-Markovianity arises from aggregating heterogeneous 
first-order Markov chains across different tactical contexts. When multiple 
formations are pooled for analysis (e.g.\ Chelsea FCW aggregating 
4-1-2-2-1, 4-2-2-1-1, and 4-1-2-1-2), the non-Markovian signal therefore 
reflects genuine higher-order dependencies in the movement dynamics rather 
than a mixture-of-Markovians artifact. 

\begin{table}[h!]
  \centering
  \caption{Debiased CMI $I(v_{t+1}; v_{t-1} \mid v_t)$ per formation.
  \# Transition Triples denotes the number of $(v_{t-1}, v_t, v_{t+1})$
  triples used for estimation, pooled across all trajectories for that
  formation; a trajectory of length $L$ contributes $L-2$ such triples.
  All $p$-values are estimated via stratified permutation test ($B=500$);
  significance is denoted *** ($p < 0.01$).}
  \label{tab:cmi}
  \setlength{\tabcolsep}{10pt}
  \renewcommand{\arraystretch}{1.3}
  \begin{tabular}{llccc}
  \hline
  \textbf{Team} & \textbf{Formation} & \textbf{CMI (nats)} & \# \textbf{Transition Triples} & \textbf{Sig.} \\
  \hline
  Bayern Munich   & 3-4-2-1 & 0.0952 & 2{,}079  & *** \\
  Bayern Munich   & 3-5-1-1 & 0.1022 & 2{,}460  & *** \\
  \hline
  Atlético Madrid & 4-4-2   & 0.0775 & 21{,}767 & *** \\
  \hline
  Chelsea FCW     & 4-1-2-1-2 & 0.0578 & 3{,}215 & *** \\
  Chelsea FCW     & 4-1-2-2-1 & 0.0456 & 3{,}191 & *** \\
  Chelsea FCW     & 4-2-2-1-1 & 0.0634 & 464     & *** \\
  \hline
  \end{tabular}
  \end{table}

\section{Empirical Results on Additional Teams and Formations}
\label{sec:appendix_mega_plot_formations_logos}

\subsection{Bayern Munich (Formations 3-4-2-1,3-5-1-1)}
\begin{figure}[htbp]
  \centering
  \includegraphics[width=0.93\textwidth]{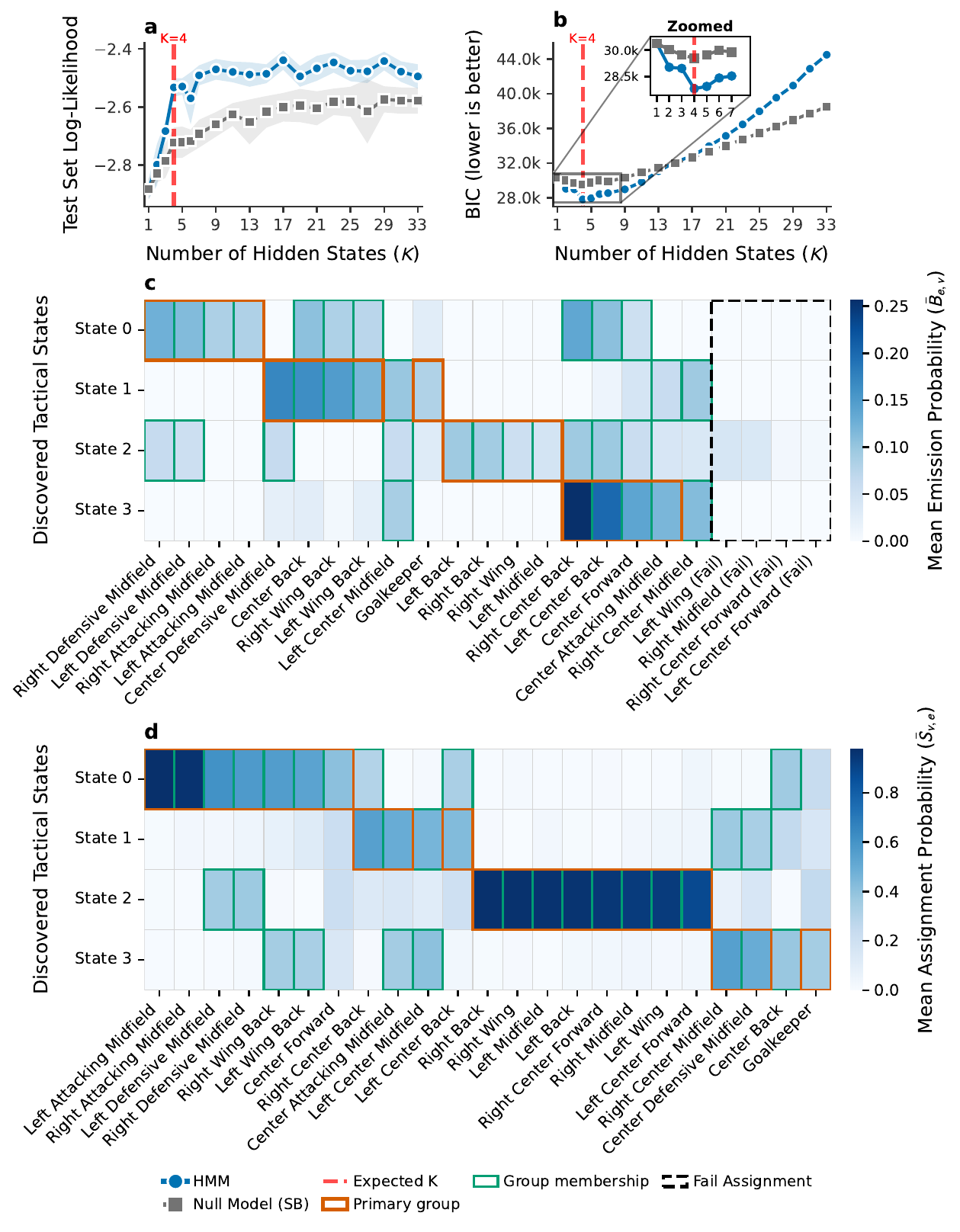}
  \caption{\textbf{Model performance and recovered tactical states for
  Bayern Munich} (formations 3-4-2-1, expected
  $K=4$ but optimal $K=9$).
  \textbf{a}, Test-set log-likelihood per step as a function of the number
  of hidden states $K$, comparing the HMM against the SB null model. The
  dashed red line marks the expected number of hidden states.
  \textbf{b}, Bayesian Information Criterion (BIC, log scale) across models
  as a function of $K$; lower is better.
  \textbf{c}, Fold-averaged emission matrix $\bar{B}_{e,v}$ at the expected
  $K$. Each role is assigned to its most probable discovered tactical
  state (highest mean emission probability across cross-validation
  folds). Significance is assessed via a standard $t$-test on the
  fold-to-fold empirical variability, at the 95\% confidence level.
  Vermillion boxes mark roles for which the top state is both
  significantly above chance and significantly more probable than the
  role's next most likely state, defining the composition of each
  recovered hyperedge. Blue boxes mark roles significantly above chance
  for a given state but without a single dominant state, indicating
  meaningful association with more than one state. Roles not significant
  above chance for any state are collected under the dashed black box
  and labelled \textit{Fail}.}
  \label{fig:bayern_munich_panel1_SI}
\end{figure}

Here we present the full results for the Bayern Munich (Supplementary Figure \ref{fig:bayern_munich_panel1_SI}) obtained from 7 matches with formation
3-4-2-1 and 3-5-1-1, comprising a total of 713 pass trajectories with a mean length of
$8.4\pm 5.8$ steps across $|V| = 23$ unique nodes (vertices). 
Note that the Jaccard similarity between the binarized group-membership
matrices $\hat{S}$ and $\hat{B}$ (based on their respective significant,
green-boxed entries) is 0.5. This indicates that the higher-order structure
is directed: there is an asymmetry between a vertex's affinity toward a
hyperedge and a hyperedge's affinity toward that vertex, analogous to a
bipartite graph with asymmetric edge weights.

\begin{figure}[htbp]
    \centering
    \includegraphics[width=0.7\textwidth]{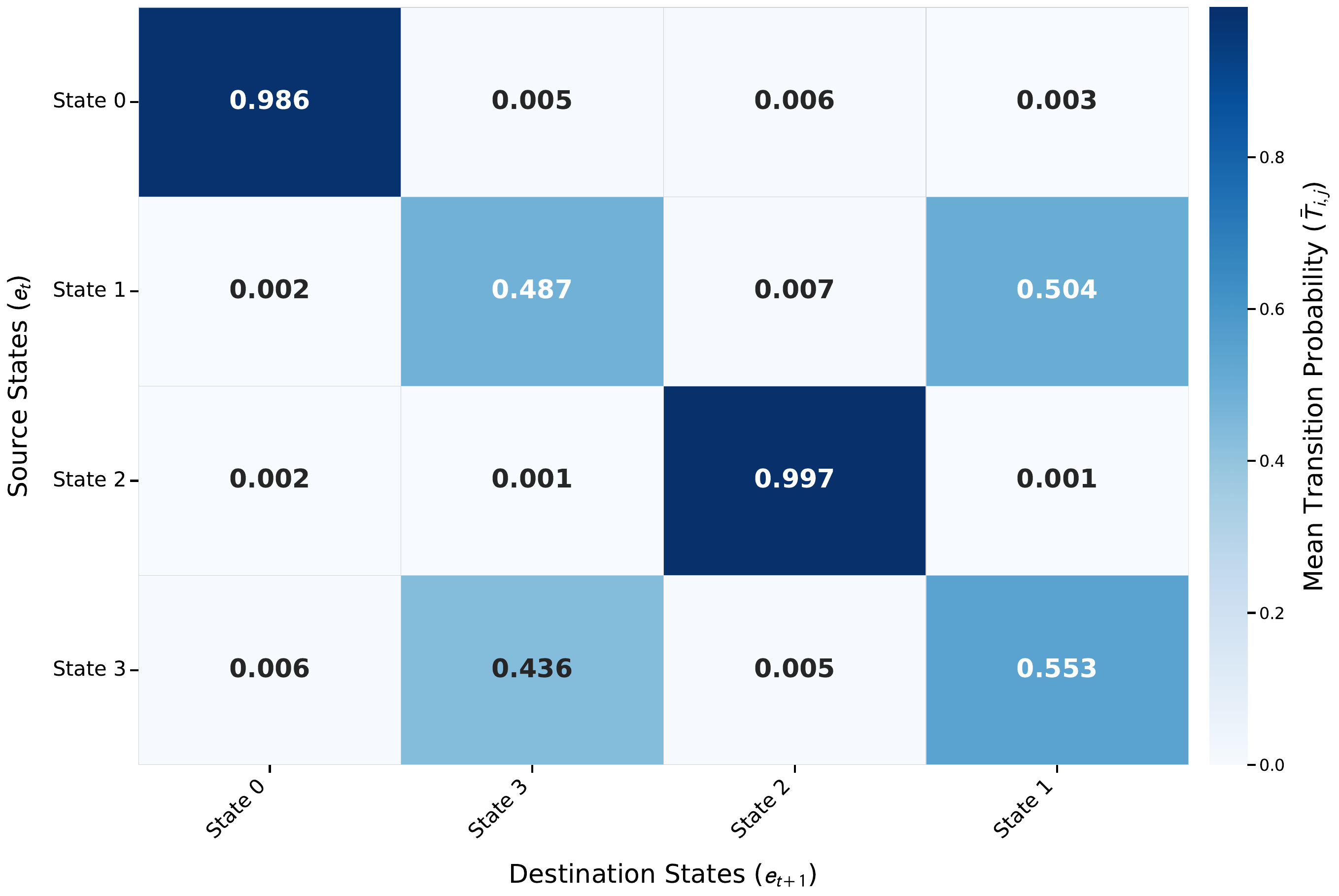}
    \caption{\textbf{Recovered transition matrix $\bar{T}_{i,j}$ for Bayern
    Munich} (formations 3-4-2-1 / 3-5-1-1). Mean transition probability
    between discovered tactical states, fold-averaged across cross-validation
    folds.}
    \label{fig:recovered_T_bayern}
\end{figure}

\clearpage
\subsection{Atlético Madrid (Formation 4-4-2)}
\begin{figure}[htbp]
  \centering
  \includegraphics[width=0.93\textwidth]{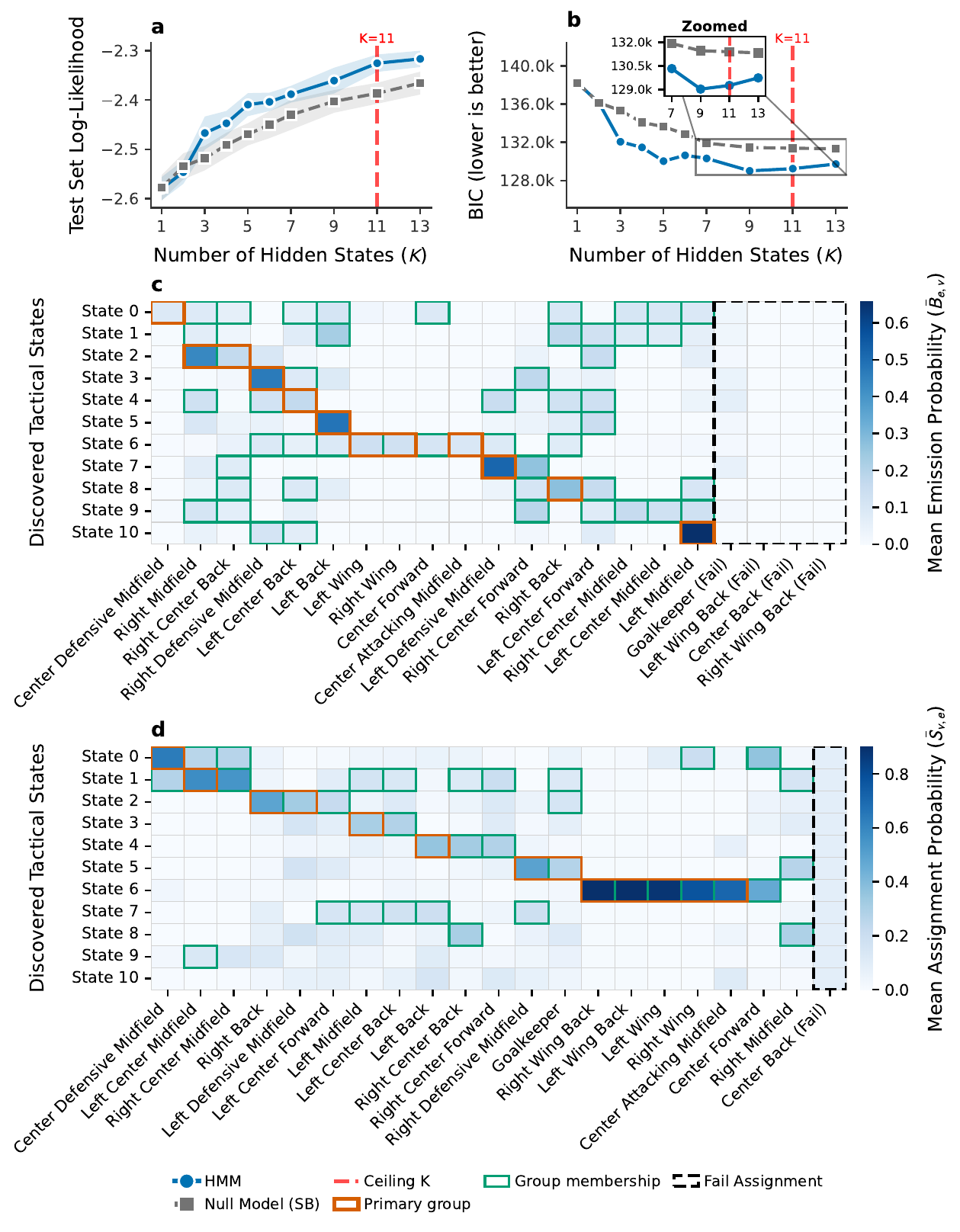}
  \caption{\textbf{Model performance and recovered tactical states for
  Atletico Madrid} (formations 4-4-2, expected
  $K=3$ but optimal $K=11$).
  \textbf{a}, Test-set log-likelihood per step as a function of the number
  of hidden states $K$, comparing the HMM against the SB null model. The
  dashed red line marks the expected number of hidden states.
  \textbf{b}, Bayesian Information Criterion (BIC, log scale) across models
  as a function of $K$; lower is better.
  \textbf{c}, Fold-averaged emission matrix $\bar{B}_{e,v}$ at the expected
  $K$. Each role is assigned to its most probable discovered tactical
  state (highest mean emission probability across cross-validation
  folds). Significance is assessed via a standard $t$-test on the
  fold-to-fold empirical variability, at the 95\% confidence level.
  Vermillion boxes mark roles for which the top state is both
  significantly above chance and significantly more probable than the
  role's next most likely state, defining the composition of each
  recovered hyperedge. Blue boxes mark roles significantly above chance
  for a given state but without a single dominant state, indicating
  meaningful association with more than one state. Roles not significant
  above chance for any state are collected under the dashed black box
  and labelled \textit{Fail}.}
  \label{fig:atletico_442_SI}
\end{figure}

Here we present the results for Altético Madrid (Supplementary Figure \ref{fig:atletico_442_SI}, Supplementary 
Table \ref{tab:model_comparison_atletico_madrid_442_k3}) obtained from 46 matches with formation
4-4-2, comprising a total of 5002 pass trajectories with a mean length of
$6.4 \pm 4.0$ steps across $|V| = 21$ unique nodes (vertices).

\begin{table}[htbp]
\centering
\begin{threeparttable}
\caption{Model comparison at expected $K$ ($K=3$) and at its largest
full-dataset BIC-dominant $K=11$. Test-set log-likelihood (mean $\pm$ 1 s.d.
across $n=10$ cross-validation folds), full-dataset BIC, and the difference in
test log-likelihood relative to MC1 ($\alpha=0.5$), the strongest
baseline.}
\label{tab:model_comparison_atletico_madrid_442_k3}
\setlength{\tabcolsep}{27pt}
\begin{tabular}{l c c c}
\toprule
\textbf{Model} & \textbf{Test LL (nats)} & \textbf{BIC} & {$\Delta$} \\
\midrule
Null Model (SB) & $-2.518 \pm 0.025$ & \num{131323} & -0.164 \\
MC2 ($\alpha=0.5$) & $-2.358 \pm 0.022$ & \num{210980} & -0.004 \\
MC1 ($\alpha=0.5$) & $-2.353 \pm 0.020$ & \num{129683} & ref. \\
\textbf{HMM ($K=3$)} & $-2.467 \pm 0.033$ & \num{132062} & -0.114 \\
HMM ($K=11$)$^\dagger$ & $\mathbf{-2.326 \pm 0.018}$ & \textbf{\num{129251}} & +0.028*** \\
\bottomrule
\end{tabular}
\begin{tablenotes}\footnotesize
\item[$\dagger$] $K=11$ is the largest $K$ at which the HMM's BIC (computed on a single fit to the full dataset) is lower than or equal to every baseline's full-dataset BIC. This is a descriptive comparison of point estimates, not a significance claim.
\item[*/**/***] $p<0.05/0.01/0.001$ (one-sided Nadeau--Bengio corrected paired $t$-test on test log-likelihood; $n=10$ folds).
\item[$\Delta$] Difference in mean test log-likelihood relative to MC1 ($\alpha=0.5$), the strongest baseline (positive favors the row's model). This is the only valid basis for a significance claim, since it is computed on paired, fold-matched differences; the mean $\pm$ 1 s.d. reported per row describes each model's own fold-to-fold variability and should not be compared for overlap. BIC values are point estimates from a single full-dataset fit and are not subject to this test.
\end{tablenotes}
\end{threeparttable}
\end{table}

\begin{figure}[htbp]
    \centering
    \includegraphics[width=0.7\textwidth]{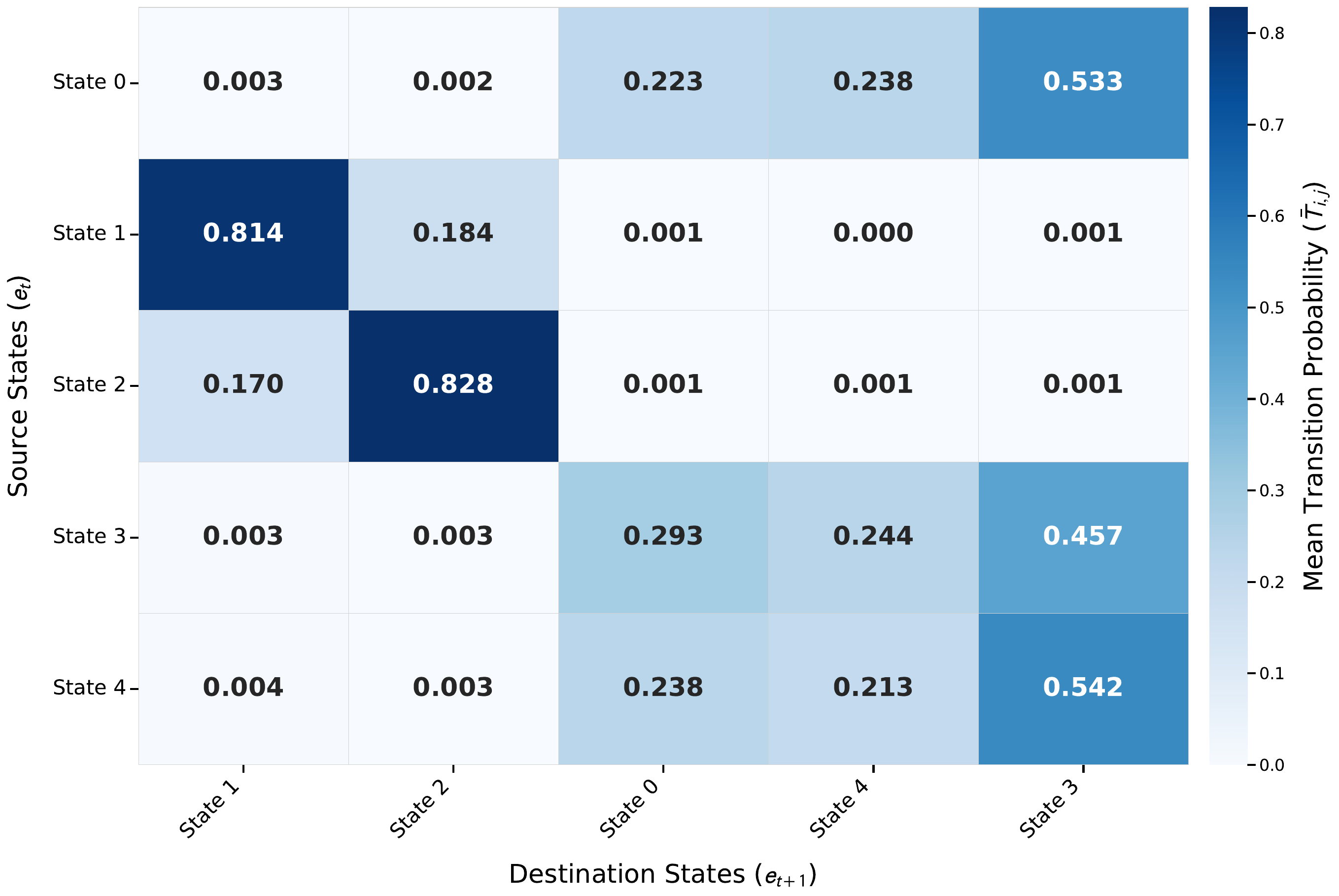}
    \caption{\textbf{Recovered transition matrix $\bar{T}_{i,j}$ for
    Atl\'etico Madrid} (formation 4-4-2). Mean transition probability
    between discovered tactical states, fold-averaged across
    cross-validation folds.}
    \label{fig:recovered_T_atletico}
\end{figure}

\clearpage
\subsection{Chelsea FCW (Formations (4-1-2-2-1 / 4-2-2-1-1 / 4-1-2-1-2))}
\begin{figure}[htbp]
  \centering
  \includegraphics[width=0.93\textwidth]{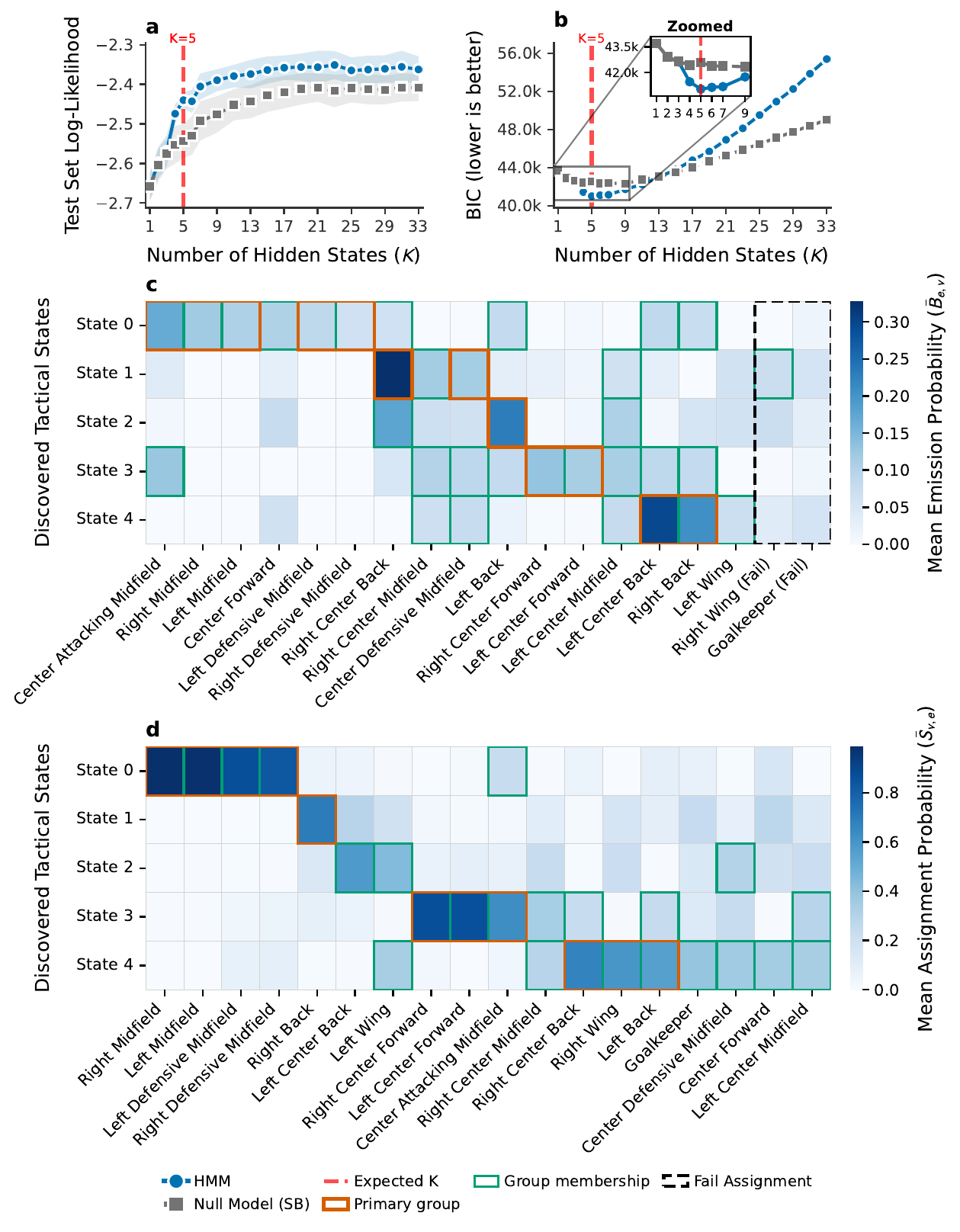}
  \caption{\textbf{Model performance and recovered tactical states for
  Chelsea FCW} (formations 4-1-2-2-1 / 4-2-2-1-1 / 4-1-2-1-2, expected
  $K=5$).
  \textbf{a}, Test-set log-likelihood per step as a function of the number
  of hidden states $K$, comparing the HMM against the SB null model. The
  dashed red line marks the expected number of hidden states.
  \textbf{b}, Bayesian Information Criterion (BIC, log scale) across models
  as a function of $K$; lower is better.
  \textbf{c}, Fold-averaged emission matrix $\bar{B}_{e,v}$ at the expected
  $K$. Each role is assigned to its most probable discovered tactical
  state (highest mean emission probability across cross-validation
  folds). Significance is assessed via a standard $t$-test on the
  fold-to-fold empirical variability, at the 95\% confidence level.
  Vermillion boxes mark roles for which the top state is both
  significantly above chance and significantly more probable than the
  role's next most likely state, defining the composition of each
  recovered hyperedge. Blue boxes mark roles significantly above chance
  for a given state but without a single dominant state, indicating
  meaningful association with more than one state. Roles not significant
  above chance for any state are collected under the dashed black box
  and labelled \textit{Fail}.}
  \label{fig:SI-chelsea}
\end{figure}

Here we present the results for Chelsea FCW (Supplementary Figure \ref{fig:SI-chelsea}, 
Supplementary Table \ref{tab:model_comparison_k5_chelsea}) obtained from 12 matches across three
formations (4-1-2-2-1, 4-2-2-1-1, and 4-1-2-1-2), comprising a total of
1332 pass trajectories with a mean length of $7.2 \pm 4.7$ steps across
$|V| = 18$ unique nodes (vertices).

\begin{table}[htbp]
  \centering
  \begin{threeparttable}
    \caption{Model comparison for Chelsea FCW (formations 4-1-2-2-1, 4-2-2-1-1, and 4-1-2-1) expected $K$ ($K=5$) and at its largest
    BIC-dominant $K=11$. Test-set log-likelihood (mean $\pm$ 1 s.d.
    across $n=10$ cross-validation folds), BIC, and the difference in
    test log-likelihood relative to MC1 ($\alpha=0.5$), the strongest
    baseline.}
  \label{tab:model_comparison_k5_chelsea}
  \setlength{\tabcolsep}{27pt}
  \begin{tabular}{l c c c}
    \toprule
    \textbf{Model} & \textbf{Test LL (nats)} & \textbf{BIC} & {$\Delta$} \\
    \midrule
    Null Model (SB) & $-2.543 \pm 0.043$ & \num{42325} & -0.141 \\
    MC2 ($\alpha=0.5$) & $-2.448 \pm 0.035$ & \num{86133} & -0.046 \\
    MC1 ($\alpha=0.5$) & $\mathbf{-2.402 \pm 0.035}$ & \num{41593} & ref. \\
    \textbf{HMM ($K=5$)} & $-2.440 \pm 0.020$ & \textbf{\num{41009}} & -0.038 \\
    HMM ($K=7$)$^\dagger$ & $-2.405 \pm 0.033$ & \num{41155} & -0.003 \\
    \bottomrule
    \end{tabular}
    \begin{tablenotes}\footnotesize
    \item[$\dagger$] $K=7$ is the largest $K$ at which the HMM's BIC (computed on a single fit to the full dataset) is lower than or equal to every baseline's full-dataset BIC. This is a descriptive comparison of point estimates, not a significance claim.
    \item[$\Delta$] Difference in mean test log-likelihood relative to MC1 ($\alpha=0.5$), the strongest baseline (positive favors the row's model). This is the only valid basis for a significance claim, since it is computed on paired, fold-matched differences; the mean $\pm$ 1 s.d. reported per row describes each model's own fold-to-fold variability and should not be compared for overlap. BIC values are point estimates from a single full-dataset fit and are not subject to this test.
    \end{tablenotes}
    \end{threeparttable}
    \end{table}

\clearpage
\subsection{Atlético Madrid (All Formations Size 3)}
\begin{figure}[htbp]
  \centering
  \includegraphics[width=0.93\textwidth]{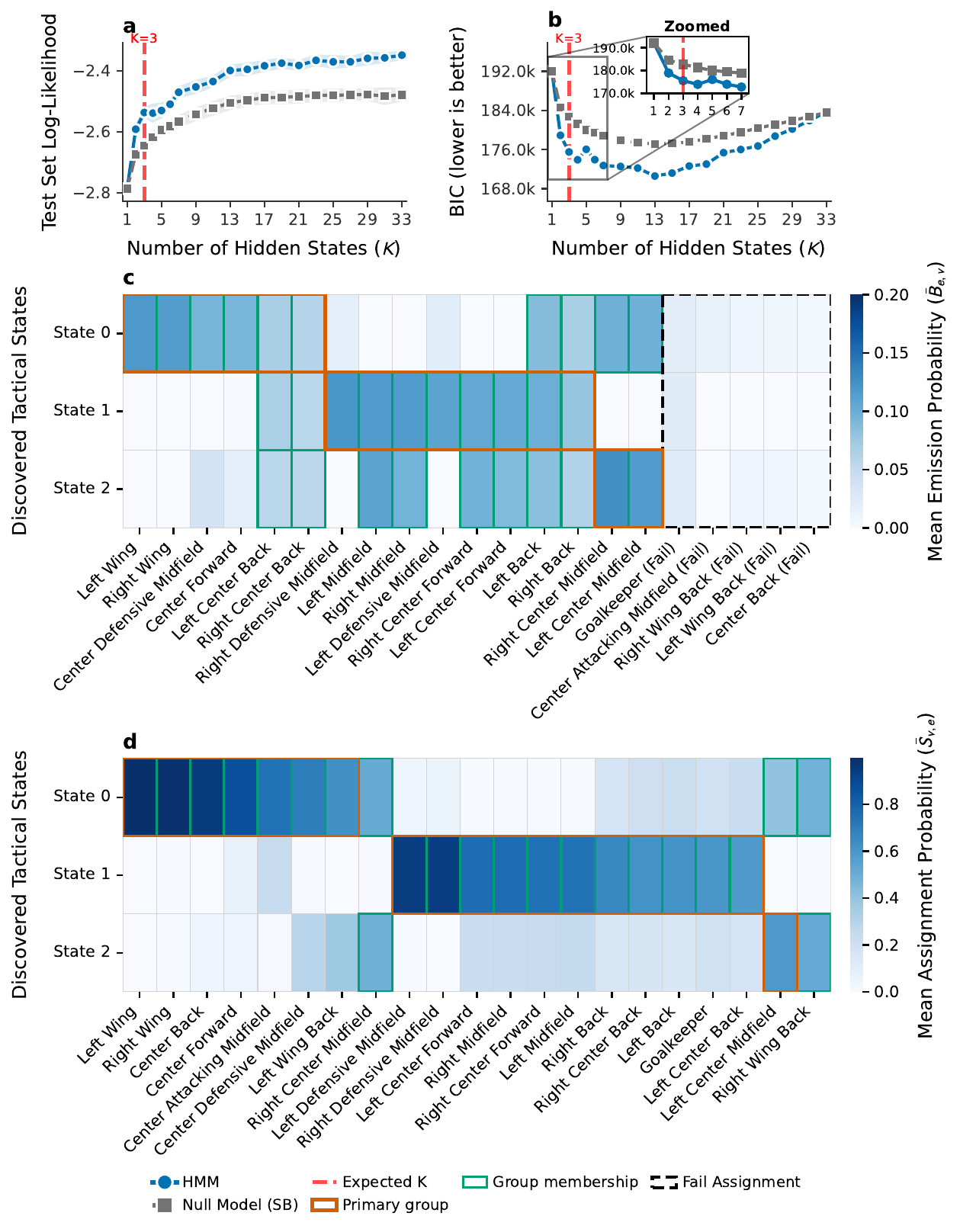}
  \caption{\textbf{Model performance and recovered tactical states for
  Atletico Madrid} (All formations size 3, expected
  $K=3$).
  \textbf{a}, Test-set log-likelihood per step as a function of the number
  of hidden states $K$, comparing the HMM against the SB null model. The
  dashed red line marks the expected number of hidden states.
  \textbf{b}, Bayesian Information Criterion (BIC, log scale) across models
  as a function of $K$; lower is better.
  \textbf{c}, Fold-averaged emission matrix $\bar{B}_{e,v}$ at the expected
  $K$. Each role is assigned to its most probable discovered tactical
  state (highest mean emission probability across cross-validation
  folds). Significance is assessed via a standard $t$-test on the
  fold-to-fold empirical variability, at the 95\% confidence level.
  Vermillion boxes mark roles for which the top state is both
  significantly above chance and significantly more probable than the
  role's next most likely state, defining the composition of each
  recovered hyperedge. Blue boxes mark roles significantly above chance
  for a given state but without a single dominant state, indicating
  meaningful association with more than one state. Roles not significant
  above chance for any state are collected under the dashed black box
  and labelled \textit{Fail}.}
  \label{fig:atletico_size_3}
\end{figure}

Here we present the results for Atlético Madrid (Supplementary Figure \ref{fig:atletico_size_3}, Supplementary Table \ref{tab:model_comparison_k3_atletico}) team obtained from 58 matches across four
formations (4-4-2, 3-4-3, 3-5-2, and 4-3-3), comprising a total of 6301
pass trajectories with a mean length of $6.5 \pm 4.2$ steps across
$|V| = 21$ unique nodes (vertices).

\begin{table}[htbp]
  \centering
  \begin{threeparttable}
  \caption{Model comparison for Altetico Madrid (all formations size 3) at expected $K$ ($K=3$) and at its largest
  BIC-dominant $K=17$. Test-set log-likelihood (mean $\pm$ 1 s.d.
  across $n=10$ cross-validation folds), BIC, and the difference in
  test log-likelihood relative to MC2 ($\alpha=0.5$), the strongest
  baseline.}
  \label{tab:model_comparison_k3_atletico}
  \setlength{\tabcolsep}{27pt}
  \begin{tabular}{l c c c}
  \toprule
  \textbf{Model} & \textbf{Test LL (nats)} & \textbf{BIC} & {$\Delta$} \\
  \midrule
  Null Model (SB) & $-2.645 \pm 0.021$ & \num{177149} & -0.241 \\
  MC1 ($\alpha=0.5$) & $-2.477 \pm 0.021$ & \num{174120} & -0.073 \\
  MC2 ($\alpha=0.5$) & $-2.404 \pm 0.020$ & \num{252167} & ref. \\
  \textbf{HMM ($K=3$)} & $-2.537 \pm 0.018$ & \num{175511} & -0.133 \\
  HMM ($K=19$)$^\dagger$ & $\mathbf{-2.375 \pm 0.018}$ & \textbf{\num{173032}} & +0.030** \\
  \bottomrule
  \end{tabular}
  \begin{tablenotes}\footnotesize
  \item[$\dagger$] $K=19$ is the largest $K$ at which the HMM's BIC (computed on a single fit to the full dataset) is lower than or equal to every baseline's full-dataset BIC. This is a descriptive comparison of point estimates, not a significance claim.
  \item[*/**/***] $p<0.05/0.01/0.001$ (one-sided Nadeau--Bengio corrected paired $t$-test on test log-likelihood; $n=10$ folds).
  \item[$\Delta$] Difference in mean test log-likelihood relative to MC2 ($\alpha=0.5$), the strongest baseline (positive favors the row's model). This is the only valid basis for a significance claim, since it is computed on paired, fold-matched differences; the mean $\pm$ 1 s.d. reported per row describes each model's own fold-to-fold variability and should not be compared for overlap. BIC values are point estimates from a single full-dataset fit and are not subject to this test.
  \end{tablenotes}
  \end{threeparttable}
  \end{table}

\clearpage
\subsection{Bayern Munich (All Formations Size 3)}
\begin{figure}[htbp]
  \centering
  \includegraphics[width=0.93\textwidth]{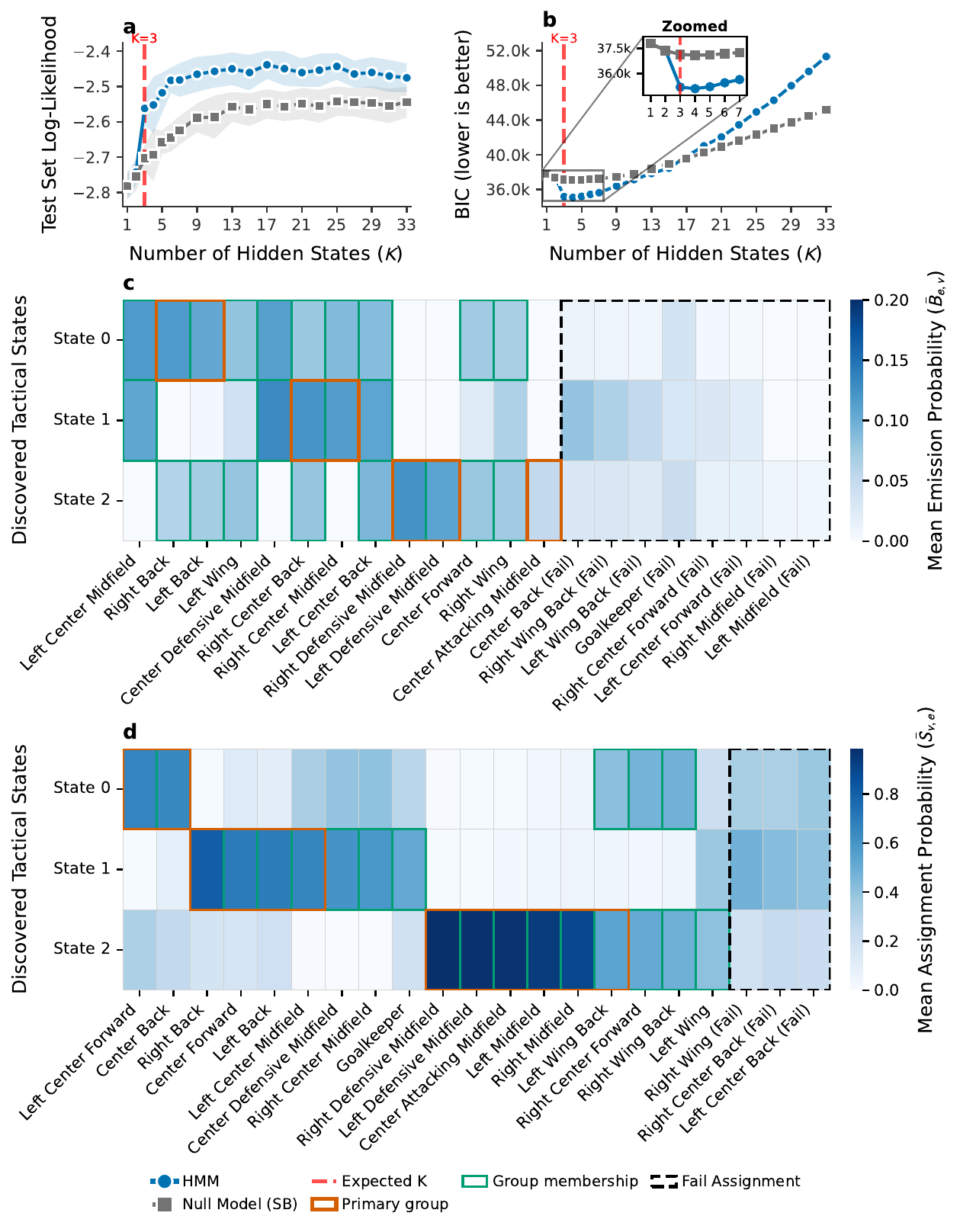}
  \caption{\textbf{Model performance and recovered tactical states for
  Bayern Munich} (All formations size 3, expected
  $K=3$).
  \textbf{a}, Test-set log-likelihood per step as a function of the number
  of hidden states $K$, comparing the HMM against the SB null model. The
  dashed red line marks the expected number of hidden states.
  \textbf{b}, Bayesian Information Criterion (BIC, log scale) across models
  as a function of $K$; lower is better.
  \textbf{c}, Fold-averaged emission matrix $\bar{B}_{e,v}$ at the expected
  $K$. Each role is assigned to its most probable discovered tactical
  state (highest mean emission probability across cross-validation
  folds). Significance is assessed via a standard $t$-test on the
  fold-to-fold empirical variability, at the 95\% confidence level.
  Vermillion boxes mark roles for which the top state is both
  significantly above chance and significantly more probable than the
  role's next most likely state, defining the composition of each
  recovered hyperedge. Blue boxes mark roles significantly above chance
  for a given state but without a single dominant state, indicating
  meaningful association with more than one state. Roles not significant
  above chance for any state are collected under the dashed black box
  and labelled \textit{Fail}.}
  \label{fig:bayern_size_3}
\end{figure}

Here we present the results for Bayern Munich (Supplementary Figure \ref{fig:bayern_size_3}, Supplementary Table \ref{tab:model_comparison_k3_bayern}) obtained from 9 matches across three
formations (3-5-2, 4-3-3, and 3-4-3), comprising a total of 940 pass
trajectories with a mean length of $8.2 \pm 5.8$ steps across $|V| = 21$
unique nodes (vertices).

\begin{table}[htbp]
  \centering
  \begin{threeparttable}
  \caption{Model comparison for Bayern Munich (all formations size 3) at expected $K$ ($K=3$) and at its largest
  BIC-dominant $K=9$. Test-set log-likelihood (mean $\pm$ 1 s.d.
  across $n=10$ cross-validation folds), BIC, and the difference in
  test log-likelihood relative to MC1 ($\alpha=0.5$), the strongest
  baseline.}
  \label{tab:model_comparison_k3_bayern}
  \setlength{\tabcolsep}{27pt}
  \begin{tabular}{l c c c}
    \toprule
    \textbf{Model} & \textbf{Test LL (nats)} & \textbf{BIC} & {$\Delta$} \\
    \midrule
    Null Model (SB) & $-2.703 \pm 0.037$ & \num{37105} & -0.172 \\
    MC2 ($\alpha=0.5$) & $-2.607 \pm 0.049$ & \num{108834} & -0.077 \\
    MC1 ($\alpha=0.5$) & $-2.531 \pm 0.038$ & \num{37230} & ref. \\
    \textbf{HMM ($K=3$)} & $-2.562 \pm 0.048$ & \textbf{\num{35190}} & -0.031 \\
    HMM ($K=9$)$^\dagger$ & $\mathbf{-2.465 \pm 0.050}$ & \num{36415} & +0.066** \\
    \bottomrule
    \end{tabular}
    \begin{tablenotes}\footnotesize
    \item[$\dagger$] $K=9$ is the largest $K$ at which the HMM's BIC (computed on a single fit to the full dataset) is lower than or equal to every baseline's full-dataset BIC. This is a descriptive comparison of point estimates, not a significance claim.
    \item[*/**/***] $p<0.05/0.01/0.001$ (one-sided Nadeau--Bengio corrected paired $t$-test on test log-likelihood; $n=10$ folds).
    \item[$\Delta$] Difference in mean test log-likelihood relative to MC1 ($\alpha=0.5$), the strongest baseline (positive favors the row's model). This is the only valid basis for a significance claim, since it is computed on paired, fold-matched differences; the mean $\pm$ 1 s.d. reported per row describes each model's own fold-to-fold variability and should not be compared for overlap. BIC values are point estimates from a single full-dataset fit and are not subject to this test.
    \end{tablenotes}
    \end{threeparttable}
    \end{table}

\clearpage
\subsection{Chelsea FCW (All Formations Size 3)}
\label{sec:supp-Chelsea_FCW_size_3}
\begin{figure}[htbp]
  \centering
  \includegraphics[width=0.93\textwidth]{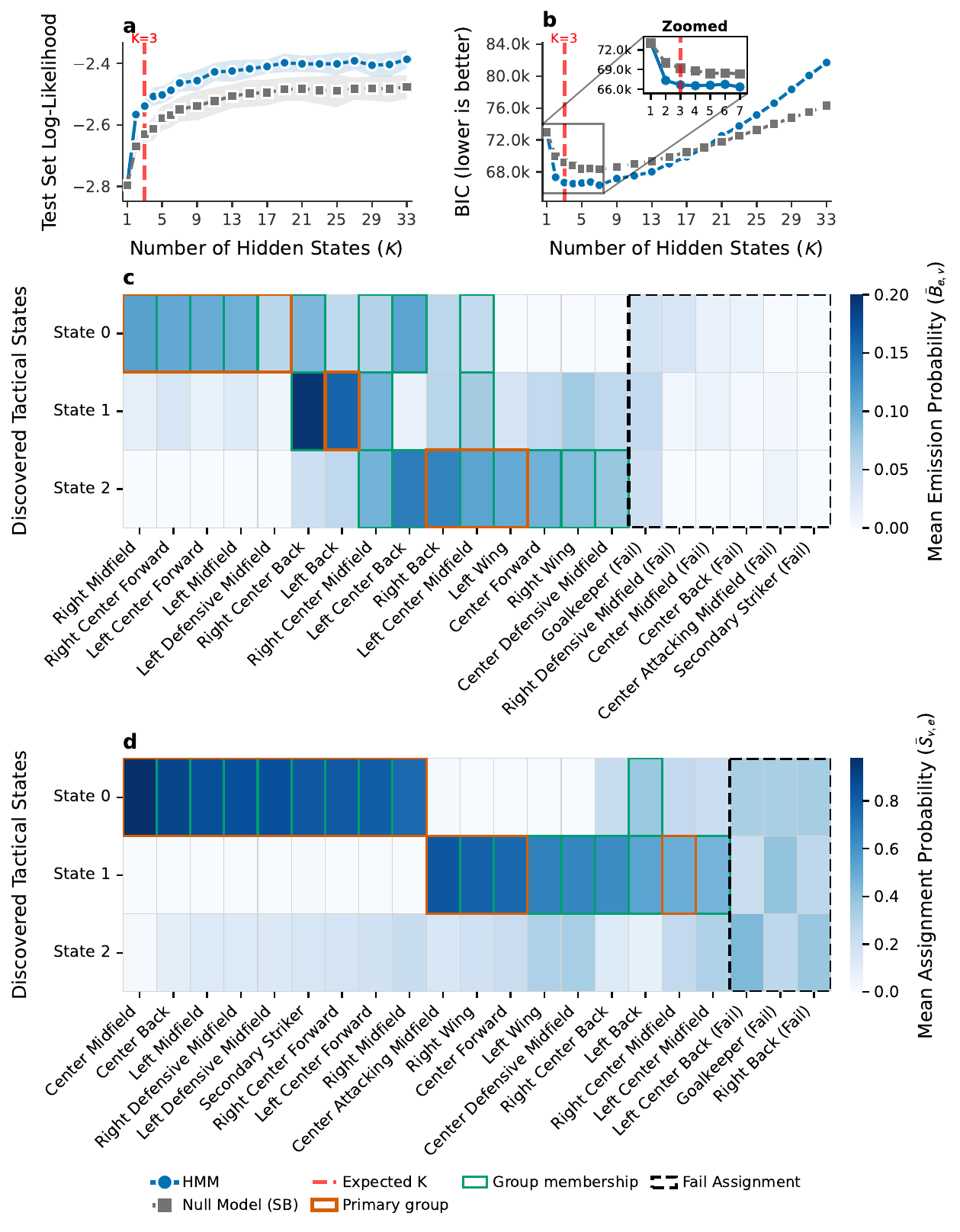}
  \caption{\textbf{Model performance and recovered tactical states for
  Chelsea FCW} (All formations size 3, expected
  $K=3$).
  \textbf{a}, Test-set log-likelihood per step as a function of the number
  of hidden states $K$, comparing the HMM against the SB null model. The
  dashed red line marks the expected number of hidden states.
  \textbf{b}, Bayesian Information Criterion (BIC, log scale) across models
  as a function of $K$; lower is better.
  \textbf{c}, Fold-averaged emission matrix $\bar{B}_{e,v}$ at the expected
  $K$. Each role is assigned to its most probable discovered tactical
  state (highest mean emission probability across cross-validation
  folds). Significance is assessed via a standard $t$-test on the
  fold-to-fold empirical variability, at the 95\% confidence level.
  Vermillion boxes mark roles for which the top state is both
  significantly above chance and significantly more probable than the
  role's next most likely state, defining the composition of each
  recovered hyperedge. Blue boxes mark roles significantly above chance
  for a given state but without a single dominant state, indicating
  meaningful association with more than one state. Roles not significant
  above chance for any state are collected under the dashed black box
  and labelled \textit{Fail}.}
  \label{fig:chelsea_size_3}
\end{figure}

Here we present the results Chelsea FCW (Supplementary Figure \ref{fig:chelsea_size_3}, 
Supplementary Table \ref{tab:model_comparison_k3_chelsea}) obtained from 20 matches across three
formations (4-4-2, 4-3-3, and 3-5-2), comprising a total of 2235 pass
trajectories with a mean length of $6.8 \pm 4.3$ steps across $|V| = 21$
unique nodes (vertices).

\begin{table}[htbp]
  \centering
  \begin{threeparttable}
  \caption{Model comparison for Chelsea FCW (all formations size 3) at expected $K$ ($K=3$) and at its largest
  BIC-dominant $K=9$. Test-set log-likelihood (mean $\pm$ 1 s.d.
  across $n=10$ cross-validation folds), BIC, and the difference in
  test log-likelihood relative to MC1 ($\alpha=0.5$), the strongest
  baseline.}
  \label{tab:model_comparison_k3_chelsea}
  \setlength{\tabcolsep}{27pt}
  \begin{tabular}{l c c c}
    \toprule
    \textbf{Model} & \textbf{Test LL (nats)} & \textbf{BIC} & {$\Delta$} \\
    \midrule
    Null Model (SB) & $-2.630 \pm 0.028$ & \num{68358} & -0.155 \\
    MC2 ($\alpha=0.5$) & $-2.484 \pm 0.037$ & \num{143382} & -0.009 \\
    MC1 ($\alpha=0.5$) & $-2.475 \pm 0.034$ & \num{67770} & ref. \\
    \textbf{HMM ($K=3$)} & $-2.539 \pm 0.021$ & \textbf{\num{66678}} & -0.064 \\
    HMM ($K=11$)$^\dagger$ & $\mathbf{-2.428 \pm 0.029}$ & \num{67524} & +0.047*** \\
    \bottomrule
    \end{tabular}
    \begin{tablenotes}\footnotesize
    \item[$\dagger$] $K=11$ is the largest $K$ at which the HMM's BIC (computed on a single fit to the full dataset) is lower than or equal to every baseline's full-dataset BIC. This is a descriptive comparison of point estimates, not a significance claim.
    \item[*/**/***] $p<0.05/0.01/0.001$ (one-sided Nadeau--Bengio corrected paired $t$-test on test log-likelihood; $n=10$ folds).
    \item[$\Delta$] Difference in mean test log-likelihood relative to MC1 ($\alpha=0.5$), the strongest baseline (positive favors the row's model). This is the only valid basis for a significance claim, since it is computed on paired, fold-matched differences; the mean $\pm$ 1 s.d. reported per row describes each model's own fold-to-fold variability and should not be compared for overlap. BIC values are point estimates from a single full-dataset fit and are not subject to this test.
    \end{tablenotes}
    \end{threeparttable}
    \end{table}

\newpage

\section{Multi Seeding Results on Football Data}
\label{sec:appendix_mega_plot_seeds_logos}

To check that the results of the main text do not depend on a single random draw (EM initialisation, train/test and cross-validation fold partitions), we repeat the full pipeline over $n=9$ independent seeds for Bayern Munich (3-4-2-1, 3-5-1-1; Supplementary Table~\ref{tab:model_comparison_multiseed_k4_bayern_munich}) and also added results for Atlético Madrid (4-4-2; Supplementary Table~\ref{tab:model_comparison_multiseed_atletico_k3}).
For Atlético Madrid, the results shown in the main text with the optimal $K=11$ not matching the expected one $K=3$ generalize across seeds.
For Bayern Munich, the HMM at the domain-expected $K=4$ already achieves the lowest BIC of any model, including its own higher-$K$ variant, and its log-likelihood is indistinguishable from MC1 ($\Delta=-0.023$, not significant). However, at each seed's ceiling $K$ ($K=9$), the HMM's predictive advantage over MC1 strengthens and becomes significant ($\Delta=+0.070$, $p<0.01$, paired one-sided Wilcoxon test), though at the cost of a higher BIC than the $K=4$ solution, reflecting the added parameter cost of the larger model. This again shows the trade-off between predictive performance and interpretability of the HMM.

\begin{table}[h!]
  \centering
  \begin{threeparttable}
  \caption{Model comparison for Atletico Madrid (4-4-2) at expected $K$ ($K=3$),
  and at each seed's largest full-dataset-BIC-dominant $K$ ($K \in [11, 13]$),
  aggregated across seeds with both cross-validation and full-dataset BIC
  results available ($n=9$). For each model, test-set
  log-likelihood is averaged within a seed across cross-validation folds and
  BIC is computed from a single fit to that seed's full dataset; the mean
  $\pm$ 1 s.d. across seeds of these per-seed values is reported, alongside
  the significance of the log-likelihood gap relative to MC1 ($\alpha=0.5$),
  the strongest baseline.}
  \label{tab:model_comparison_multiseed_atletico_k3}
  \setlength{\tabcolsep}{24pt}
  \begin{tabular}{l c c c}
  \toprule
  \textbf{Model} & \textbf{Test LL} & \textbf{BIC} & {$\Delta$} \\
  \midrule
  Null Model (SB) & $-2.513 \pm 0.004$ & \num{131198} & -0.159 \\
  MC2 ($\alpha=0.5$) & $-2.357 \pm 0.001$ & \num{210980} & -0.003 \\
  MC1 ($\alpha=0.5$) & $-2.354 \pm 0.001$ & \num{129683} & ref. \\
  HMM ($K=3$) & $-2.470 \pm 0.024$ & \num{132638} & -0.116 \\
  \textbf{HMM} ($K \in [11, 13]$)$^\dagger$ & $\mathbf{-2.324 \pm 0.007}$ & \textbf{\num{129407}} & +0.030** \\
  \bottomrule
  \end{tabular}
  \begin{tablenotes}\footnotesize
  \item[$\dagger$] Per seed, computed from a single full-dataset fit per model (no k-fold), the largest $K$ at which the HMM's BIC is simultaneously lower than or equal to every baseline's BIC (9/9 seeds satisfied this; the remaining seed(s) are excluded from this row).
  \item[*/**/***] $p<0.05/0.01/0.001$, paired one-sided wilcoxon test across seed-level means ($H_1$: the row's model outperforms MC1 ($\alpha=0.5$)).
  \item[$\Delta$] Difference in mean seed-level test log-likelihood relative to MC1 ($\alpha=0.5$), the strongest baseline (positive favors the row's model). Test LL mean $\pm$ 1 s.d. per row describes each model's own seed-to-seed variability and is not a basis for significance by itself; $\Delta$, computed on seed-matched paired differences, is. BIC is a point estimate from a single full-dataset fit per seed (mean $\pm$ 1 s.d. across seeds shown descriptively) and is not subject to this significance test.
  \end{tablenotes}
  \end{threeparttable}
\end{table}

  \begin{table}[h!]
    \centering
    \begin{threeparttable}
    \caption{Model comparison for Bayern Munich (3-4-2-1,3-5-1-1) at expected $K$ ($K=4$),
    and at each seed's largest full-dataset-BIC-dominant $K$ ($K=9$),
    aggregated across seeds with both cross-validation and full-dataset BIC
    results available ($n=9$). For each model, test-set
    log-likelihood is averaged within a seed across cross-validation folds and
    BIC is computed from a single fit to that seed's full dataset; the mean
    $\pm$ 1 s.d. across seeds of these per-seed values is reported, alongside
    the significance of the log-likelihood gap relative to MC1 ($\alpha=0.5$),
    the strongest baseline.}
    \label{tab:model_comparison_multiseed_k4_bayern_munich}
    \setlength{\tabcolsep}{24pt}
    \begin{tabular}{l c c c}
    \toprule
    \textbf{Model} & \textbf{Test LL} & \textbf{BIC} & {$\Delta$} \\
    \midrule
    Null Model (SB) & $-2.745 \pm 0.014$ & \num{29610} & -0.186 \\
    MC2 ($\alpha=0.5$) & $-2.659 \pm 0.005$ & \num{123595} & -0.100 \\
    MC1 ($\alpha=0.5$) & $-2.560 \pm 0.003$ & \num{30453} & ref. \\
    \textbf{HMM} ($K=4$) & $-2.583 \pm 0.047$ & \textbf{\num{28104}} & -0.023 \\
    HMM ($K=9$)$^\dagger$ & $\mathbf{-2.489 \pm 0.014}$ & \num{29243} & +0.070** \\
    \bottomrule
    \end{tabular}
    \begin{tablenotes}\footnotesize
    \item[$\dagger$] Per seed, computed from a single full-dataset fit per model (no k-fold), the largest $K$ at which the HMM's BIC is simultaneously lower than or equal to every baseline's BIC (9/9 seeds satisfied this; the remaining seed(s) are excluded from this row).
    \item[*/**/***] $p<0.05/0.01/0.001$, paired one-sided wilcoxon test across seed-level means ($H_1$: the row's model outperforms MC1 ($\alpha=0.5$)).
    \item[$\Delta$] Difference in mean seed-level test log-likelihood relative to MC1 ($\alpha=0.5$), the strongest baseline (positive favors the row's model). Test LL mean $\pm$ 1 s.d. per row describes each model's own seed-to-seed variability and is not a basis for significance by itself; $\Delta$, computed on seed-matched paired differences, is. BIC is a point estimate from a single full-dataset fit per seed (mean $\pm$ 1 s.d. across seeds shown descriptively) and is not subject to this significance test.
    \end{tablenotes}
    \end{threeparttable}
    \end{table}

\clearpage
\newpage

\section{Tactical Roles Dictionary}
\label{sec:appendix_tactical_roles}

The state space ${V}$ relies on the $|V|=24$ distinct tactical roles defined by the data provider specification. The complete alphabetical list of these roles and their standard abbreviations is as follows:

\begin{itemize}
    \setlength{\itemsep}{0pt}
    \item \textbf{CAM}: Center Attacking Midfield
    \item \textbf{CB}: Center Back
    \item \textbf{CDM}: Center Defensive Midfield
    \item \textbf{CF}: Center Forward
    \item \textbf{CM}: Center Midfield
    \item \textbf{GK}: Goalkeeper
    \item \textbf{LAM}: Left Attacking Midfield
    \item \textbf{LB}: Left Back
    \item \textbf{LCB}: Left Center Back
    \item \textbf{LCF}: Left Center Forward
    \item \textbf{LCM}: Left Center Midfield
    \item \textbf{LDM}: Left Defensive Midfield
    \item \textbf{LM}: Left Midfield
    \item \textbf{LW}: Left Wing
    \item \textbf{LWB}: Left Wing Back
    \item \textbf{RAM}: Right Attacking Midfield
    \item \textbf{RB}: Right Back
    \item \textbf{RCB}: Right Center Back
    \item \textbf{RCF}: Right Center Forward
    \item \textbf{RCM}: Right Center Midfield
    \item \textbf{RDM}: Right Defensive Midfield
    \item \textbf{RM}: Right Midfield
    \item \textbf{RW}: Right Wing
    \item \textbf{RWB}: Right Wing Back
\end{itemize}

\end{document}